\documentclass[conference]{IEEEtran}
\IEEEoverridecommandlockouts
\usepackage{graphicx}
\usepackage{amsmath}
\usepackage{amsfonts}
\usepackage{amssymb}
\usepackage{booktabs} 
\usepackage{subfigure}
\usepackage{psfrag}
\usepackage{color}
\usepackage{multicol}
\usepackage{tikz}
\usepackage{verbatimbox}
\usepackage{graphicx}
\usepackage{epsfig}
\usepackage{multirow}
\usepackage{lettrine}
\usepackage{cite}

\usepackage{lipsum,adjustbox}
\usetikzlibrary{calc,positioning,automata}
\usepackage{makecell}

\usepackage{algorithm}
\usepackage{algcompatible}
\usepackage{algpseudocode}

\newcommand{\bm}{\mathbf}
\newcommand{\bs}{\boldsymbol}
\newcommand{\be}{\begin{equation}}
\newcommand{\ee}{\end{equation}}
\newcommand{\bea}{\begin{eqnarray}}
\newcommand{\eea}{\end{eqnarray}}

\newcommand{\bzero}{{\bm 0}}
\newcommand{\bone}{{\bm 1}}

\newcommand{\ba}{{\bm a}}

\newcommand{\bc}{{\bm c}}

\newcommand{\e}{{\bm e}}

\newcommand{\p}{{\bm p}}

\newcommand{\br}{{\bm r}}
\newcommand{\s}{{\bm s}}
\newcommand{\x}{{\bm x}}
\newcommand{\y}{{\bm y}}
\newcommand{\z}{{\bm z}}

\newcommand{\bvar}{{\bs{\varepsilon}}}

\newcommand{\bd}{{\bf d}}

\newcommand{\bh}{{\bf h}}

\newcommand{\bA}{{\bm A}}
\newcommand{\bB}{{\bm B}}

\newcommand{\bD}{{\bm D}}
\newcommand{\bE}{{\bm E}}
\newcommand{\bF}{{\bm F}}
\newcommand{\bG}{{\bm G}}
\newcommand{\bH}{{\bm H}}
\newcommand{\bI}{{\bm I}}
\newcommand{\bJ}{{\bm J}}
\newcommand{\bO}{{\bm O}}

\newcommand{\bR}{{\bm R}}
\newcommand{\bS}{{\bm S}}

\newcommand{\bX}{{\bm X}}

\newcommand{\bZ}{{\bm Z}}

\newcommand{\I}{{\bm I}}

\newcommand{\BD}{{\boldsymbol{\mathcal D}}}

\newcommand{\bLambda}{\mbox{\boldmath$\Lambda$}}

\newcommand{\bGamma}{\mbox{\boldmath$\Gamma$}}
\newcommand{\bOmega}{\mbox{\boldmath$\Omega$}}
\newcommand{\bPsi}{\mbox{\boldmath$\Psi$}}

\newcommand{\bPhi}{\mbox{\boldmath{$\Phi$}}}

\newcommand{\bPi}{\mbox{\boldmath{$\Pi$}}}

\title{Time and Frequency Synchronization for Multiuser OTFS in Uplink\vspace{-0.1cm}}
\author{\IEEEauthorblockN{Mohsen Bayat, Sanoopkumar P.S., and Arman Farhang\vspace{-1cm}}
\thanks{
Mohsen Bayat, Sanoopkumar P.S., and Arman Farhang are with the Department of Electronic and Electrical Engineering, Trinity College Dublin, Dublin 2, Ireland
(Email: \{bayatm, pungayis, arman.farhang\}@tcd.ie).
A minor part of this submission, \cite{Bayat2025}, was presented at the IEEE WCNC 2025.
This publication has emanated from research conducted with the financial support of Taighde Éireann – Research Ireland under Grant number 19/FFP/7005(T) and 21/US/3757.
}
}

\begin{document}
\maketitle
%%%%%%%%%%%%%%%%%%%%%%%%%%%%%%%%%%%%%%%%
\begin{abstract}
In this paper, we propose time and frequency synchronization techniques for uplink multiuser orthogonal time frequency space (MU-OTFS) systems in high-mobility scenarios. This work focuses on accurately estimating and correcting timing offsets (TOs) and carrier frequency offsets (CFOs). Specifically, TO estimation is essential for locating users' pilots on the delay-time plane, while CFO estimation enhances channel estimation accuracy.
First, we propose a TO estimation technique for an existing multiuser pilot structure in MU-OTFS. We replace the impulse pilot (IMP) in this pilot structure with a more practical pilot with a cyclic prefix (PCP), referred to as single-user-inspired PCP (SU-PCP). This structure employs different Zadoff-Chu (ZC) sequences, which enables pilot separation via correlation at the receiver side. Consequently, we introduce a correlation-based TO estimation technique for uplink MU-OTFS using this pilot structure.
Next, a spectrally efficient and practical pilot pattern is proposed, where each user transmits a PCP within a shared pilot region on the delay-Doppler plane, referred to as multiuser PCP (MU-PCP). At the receiver, the second TO estimation technique utilizes a bank of filters to separate different users' signals and accurately estimate their TOs. Then, we derive a mathematical threshold range to enhance TO estimation accuracy by finding the first major peak in the correlation function rather than relying solely on the highest peak.
After locating the received users' pilot signals using one of the proposed TO estimation techniques, our proposed CFO estimation technique reduces the multi-dimensional maximum likelihood (ML) search problem into multiple one-dimensional search problems. In this technique, we apply the Chebyshev polynomials of the first kind basis expansion model (CPF-BEM) to effectively handle the time-variations of the channel in obtaining the CFO estimates for all the users. Finally, we numerically investigate the error performance of our proposed synchronization technique in high mobility scenarios for the MU-OTFS uplink. Our simulation results confirm the efficacy of the proposed technique in estimating the TOs and CFOs which also leads to an improved channel estimation performance.
\end{abstract}

%%%%%%%%%%%%%%%%%%%%%%%%%%%%%%%%%%%%%%%%
% \vspace{-0.2cm}
% \begin{IEEEkeywords}
% Delay-Doppler, Multiuser OTFS, Multiuser Pilot Structure, Synchronization, Maximum-Likelihood Estimation.
% \end{IEEEkeywords}
%%%%%%%%%%%%%%%%%%%%%%%%%%%%%%%%%%%%%%%%
\vspace{-0.2cm}
\section{introduction}\label{sec:Introduction}
\vspace{-0.1cm} 
The multipath propagation and Doppler shifts in wireless channels result in time dispersion and frequency dispersion, respectively, making these channels doubly-dispersive \cite{Jakes1974}. Waveform design seeks to enhance signal robustness against these dispersions. While orthogonal frequency division multiplexing (OFDM), with its capability to mitigate inter-symbol intereference (ISI) \cite{Gold2005}, has been the dominant waveform in the last two generations of wireless systems, it struggles to address the challenges posed by highly time-varying wireless channels that cause inter-carrier interference (ICI) \cite{Wang2006}. To effectively tackle both ISI and ICI in doubly-dispersive channels, delay-Doppler multiplexing techniques were recently proposed as a candidate technology to effectively handle time-varying channels \cite{Wei2021}.
Orthogonal time frequency space modulation (OTFS) operates in the delay-Doppler domain where the wireless channel has a sparse and time-invariant representation \cite{Wei2021}.
Since its emergence, there has been growing interest among researchers in this technology, as highlighted by the formation of a large body of literature in this area \cite{Raviteja2018}.
However, OTFS is still in its early stages of development as the majority of its literature is focused on single-user systems. 
This is while multiuser capabilities are important aspects of the future wireless networks \cite{Wei2021}, particularly concerning practical challenges in multiuser OTFS (MU-OTFS), such as synchronization.

%%%%%%%%%
Timing errors cause interference between orthogonal frequency division multiple access (OFDMA) frames and among users, leading to severe error rate degradation if not properly corrected. Additionally, inaccurate carrier frequency offset (CFO) compensation disrupts subcarrier orthogonality, which results in ICI and, consequently, multiuser interference (MUI). While subcarrier orthogonality in the time-frequency domain enables interference-free multiuser communication in this system, synchronization errors lead to significant challenges. Similar to OFDMA, MU-OTFS is highly sensitive to timing offsets (TOs) and CFOs, which can disrupt alignment between the transmitted and received signals \cite{Aminjavaheri2015, Das2021}.
One solution to this challenge in MU-OTFS is to employ a large guard interval as part of the cyclic prefix (CP) to mitigate timing errors, while also considering CFO as an additional frequency shift to the Doppler shift and Doppler spread during the channel estimation process \cite{Das2021}. As it will be shown later in this paper, estimating CFO as part of the channel reduces estimation accuracy compared to first compensating CFO and then performing channel estimation. Hence, this approach does not fully resolve synchronization challenges in MU-OTFS. Additionally, accurate TO compensation is essential for locating the users' pilots in the delay-Doppler domain. 

%%%
After signal transmission, the received signal must be synchronized at the initial stage of the receiver. This process is typically achieved through a three-phase approach in current communication standards. The first phase occurs during downlink transmission, where each  mobile terminal (MT) estimates TO and CFO using a pilot signal sent by the base station (BS). These estimated parameters are used by the MTs not only to detect the downlink data but also as reference points for synchronizing the uplink transmission. However, Doppler effects and propagation delays can cause the signals arriving at the BS to become outdated which results in residual synchronization errors in the uplink signals at the BS. Thus, the second phase involves estimating the TO and CFO in the uplink. In the final phase, the BS applies necessary corrections to these offsets and synchronizes the received signal \cite{Morelli2007}.

The existing literature on OTFS synchronization is focused on the first stage for single-user scenarios \cite{Saquib2021, Saquib2024, Chung2024, Xu2023, Sun2024, SLi2024, Bayat2022, Bayat2023}, where TO and CFO are estimated and pre-compensated or absorbed to the channel estimation stage.
In \cite{Das2021}, TO and CFO are estimated as a part of the channel in single-user scenario. However, this approach is not applicable to the uplink as accurate estimation of multiple TOs is required to locate the users' pilots. Moreover, as we show in this paper, absorbing the CFOs into the channel leads to channel estimation performance degradation. To the best of our knowledge, these problems have not been reported in the literature to date. This scenario is roughly similar to the synchronization techniques proposed for OFDMA in \cite{Morelli2006, Morelli2007_2}.
As the only work on the uplink MU-OTFS, the authors in \cite{Sinha2020} propose a closed-loop TO estimation technique, where the estimated TO is fed back to the MT. However, such an approach may lead to outdated TO estimates in the next uplink transmission. Thus, an open-loop solution can effectively address this issue. 
Consequently, the second stage is required to focus on estimating the TO and CFO in the uplink, followed by the BS applying necessary corrections to these offsets in the third synchronization stage to restore signal alignment \cite{Morelli2007}. At this stage, if one user's CFO or TO is corrected, it misaligns other users' signals. Therefore, in addition to their estimation, TO and CFO compensation itself presents another challenge for uplink transmission.

%%%%%%%%
\begin{table}[t]
\centering
\caption{Summary of Notation \vspace{-0.15cm}}
\label{tab:notation}
\begin{tabular}{ll}
\toprule
\textbf{Symbol} & \textbf{Description} \\
\midrule
$a$ & Scalar \\
[1.0ex]
$\bm{a}$ & Vector (bold lowercase) \\
[1.0ex]
$\bm{A}$ & Matrix (bold uppercase) \\
[1.0ex]
$\mathbb{C}^{M \times N}$ & Set of $M \times N$ complex matrices \\
[1.0ex]
$\bm{I}_N$ & $N \times N$ identity matrix \\
[1.0ex]
$\bm{1}_N$ & $N \times N$ all-one matrix \\
[1.0ex]
$\bm{0}_N$ & $N \times N$ all-zero matrix \\
[1.0ex]
$\bm{F}_N$ & \makecell[l]{Normalized \(N\)-point DFT matrix with elements \\ \(F[l,n]\!=\!\frac{1}{\sqrt{N}} e^{-j \frac{2 \pi l n}{N}}\), for \(l,n\!=\!0, \ldots, N\!-\!1\)} \\
[3.0ex]
$\bm{A} = {\rm circ}\{\bm{a}\}$ & Circulant matrix with first column $\bm{a}$ \\
[1.0ex]
$\bm{A}_{l'}[l,n]$ & \makecell[l]{Element \([l,n]\) of matrix \(\bm{A}\), upward shifted by \\ \(l'\) along the \(l\)-dimension} \\
[2.0ex]
$(\cdot)^{\rm H}$ & Hermitian (conjugate transpose) \\
[1.0ex]
$(\cdot)^{\rm T}$ & Transpose operator \\
[1.0ex]
$(\cdot)^{-1}$ & Matrix inverse \\
[1.0ex]
$\odot$ & Element-wise (Hadamard) product \\
[1.0ex]
$\otimes$ & Kronecker product \\
[1.0ex]
$(\!(\cdot)\!)_i$ & Circular shift of \(i\) samples\\
[1.0ex]
$(\cdot)_i$ & Modulo \(i\) operation \\
[1.0ex]
${\rm vec}\{\cdot\}$ & Vectorization of a matrix \\
[1.0ex]
${\rm diag}[\cdot]$ & Diagonal matrix formed from a vector \\
[1.0ex]
$\max_i \{ \cdot \}$ & Maximum over index \(i\) \\
[1.0ex]
$\lfloor a \rfloor$ & The greatest integer less than or equal to $a$ \\
[1.0ex]
$\delta[\cdot]$ & Discrete Dirac delta function \\
[1.0ex]
$\mathbb{A} \!=\! \{ a_0, \ldots, a_{N-1} \}$ & Set $\mathbb{A}$ including members $a_0, \ldots, a_{\!N\!-\!1}$ \\
[1.0ex]
$\emptyset$ & A set containing no elements (empty set) \\
[1.0ex]
$\mathbb{A} \cap \mathbb{B}$ & The intersection of two sets $\mathbb{A}$ and $\mathbb{B}$\\
[1.0ex]
$\mathbb{A} \cup \mathbb{B}$ & The union of two sets $\mathbb{A}$ and $\mathbb{B}$\\
[1.0ex]
$\bigcup_{i=\!1}^a \! \mathbb{A}^i$ & \makecell[l]{Union of user sets $\mathbb{A}^i$ forming $\{0,\ldots,N-1\}$} \\
\bottomrule
\end{tabular}
\vspace{-0.7cm}
\end{table}
%%%%%%%%

To address the above challenges, this paper proposes a novel synchronization technique addressing the second and third stages of the synchronization process for MU-OTFS uplink transmissions under a linear time-varying (LTV) channel. Since uplink signals are affected by users’ TOs and CFOs, aligning one user’s signal in time and frequency often misaligns other users' signals. To address this, the compound channel matrix is used to jointly estimate and compensate for all users' CFOs, while individual TO estimation for each user helps to locate the correct pilot regions.
Then, a TO estimation technique is proposed for an existing multiuser pilot structure, named single-user-inspired pilot with a cyclic prefix (SU-PCP) \cite{Raviteja2018}, which inserts PCP for different users isolated along the delay dimension. Our proposed TO estimation technique leverages the lack of correlation between different Zadoff-Chu (ZC) sequences to separate users' pilot regions in the correlation functions at the receiver of an uplink MU-OTFS system.
To prevent spectral efficiency loss, a spectrally efficient pilot pattern is proposed, where each user transmits a pilot within a shared pilot region on the delay-Doppler plane. This pilot structure is referred to as multiuser PCP (MU-PCP). At the receiver, user signals are separated using a bank of filters, followed by TO and CFO estimation through correlation-based and maximum likelihood (ML)-based approaches, respectively. The TO estimation uses a threshold to identify the first major peak of the correlation function, achieving an accuracy improvement of two orders of magnitude over using the maximum peak for TO estimation. This threshold-based method is further enhanced by mathematically deriving a range for selecting the threshold.
Additionally, an ML estimation technique is developed for frequency synchronization. By separating different users' pilots, the multidimensional search for estimating multiple CFOs is reduced to multiple line search problems. The ML technique employs the Chebyshev polynomials of the first kind basis expansion
model (CPF-BEM) to absorb the time variations of the channel coefficients into basis functions and effectively estimate the CFOs.
%%%%%%%
The key contributions of this paper are summarized as follows:
\begin{itemize}   
\item For asynchronous MU-OTFS uplink, a compound channel model is derived that incorporates the effects of TOs and CFOs in the delay-Doppler domain.
\item A novel pilot pattern for MU-OTFS, called MU-PCP, is proposed to allow all users to share a common pilot region.
\item A filter bank is introduced to separate user signals for TO estimation and to simplify ML-based CFO estimation by converting a multidimensional search problem into multiple one-dimensional search problems.
\item Novel time and frequency synchronization techniques are developed for the uplink of MU-OTFS.
\begin{itemize}
    \item A correlation-based TO estimation method is proposed, in which a threshold is employed to enhance the accuracy of the TO estimate.
    \item The threshold range is mathematically derived and used to identify the first major peak of the correlation function, yielding a highly accurate TO estimate.
    \item An ML estimation method is proposed for frequency synchronization in MU-OTFS uplink systems. The approach employs CPF-BEM to separate the time-varying components of the channel from its time-invariant part. This decomposition enables the application of ML estimation to the time-invariant component, while the known time-varying part is absorbed into the pilot term.
\end{itemize}
\end{itemize}
%%%%%%%

%%%%%%%%%
This paper is structured as follows.
Section~\ref{sec:System} presents the system model for an asynchronous MU-OTFS system in the uplink. 
In Section~\ref{sec:Ch}, the channel effect in an asynchronous MU-OTFS system, in which both TO and CFO are incorporated is derived.
Section~\ref{sec:TO} presents TO estimation techniques for the uplink MU-OTFS system.
Specifically, the proposed TO estimation method in Subsection~\ref{sec:TO1} adapts a single-user synchronization technique from OTFS to MU-OTFS, using a pilot pattern similar to the one in \cite{Raviteja2018}.
Section~\ref{sec:TO2} details the proposed pilot structure and TO estimation technique for this configuration which improved spectral efficiency.
The correlation-based TO estimation method employs a threshold to detect the first major peak of the correlation function.
This threshold range is mathematically derived in Subsection~\ref{sec:TO_err}.
Additionally, Subsection~\ref{sec:ML} introduces an ML-based technique for CFO estimation, employing a CPF-BEM model to capture time-varying channel effects into the pilot matrix. Section~\ref{sec:BW} analyzes spectral efficiency and computational complexity to assess the effectiveness of the proposed pilot structure and synchronization techniques. 
Section~\ref{sec:Result} evaluates the performance of the proposed techniques through simulations. Finally, Section~\ref{sec:Conclusion} concludes the paper.

\vspace{-0.15cm}
\section{System Model} \label{sec:System}
\vspace{-0.05cm}
%%%%%%%%%%%%%%%%
In this paper, we consider an uplink OTFS system where $Q$ users are simultaneously communicating with the BS simultaneously. At each MT, the information bits are mapped onto quadrature amplitude modulation (QAM) constellation and placed on a delay-Doppler grid with $M$ delay bins and $N$ Doppler bins.
Given the delay spacing of $\Delta \tau$, each delay block with $M$ delay bins has a duration of $T=M\Delta \tau$, and the total frame duration is $T_{\rm{f}} = NT = MN \Delta \tau$. Hence, the Doppler spacing is $\Delta \nu=\frac{1}{T_{\rm{f}}}=\frac{1}{NT}=\frac{1}{M N \Delta \tau}$.

We present the sets of $M_q$ delay bins and $N_q$ Doppler bins allocated to the users $q=0,\ldots,Q-1$ as $\mathbb{U}_\tau^q$ and $\mathbb{U}_\nu^q$, respectively. The delay-Doppler resources allocated to any pair of distinct users $q,p\in\{1,\ldots,Q \}$ belong to mutually exclusive sets $\mathbb{U}_{\tau/\nu}^q \cap \mathbb{U}_{\tau/\nu}^p=\emptyset$ where $\bigcup_{q=1}^Q \mathbb{U}_\tau^q=\{0,\ldots,M-1\}$ and $\bigcup_{q=1}^Q \mathbb{U}_\nu^q=\{0,\ldots,N-1\}$.
After stacking the QAM symbols of user $q$ into $M_q\times N_q$ data matrix $\BD^q$, they are mapped onto their corresponding delay-Doppler bins. This is done by the delay and Doppler resource allocation matrices $\bGamma_\tau^q$ and $\bGamma_\nu^q$, respectively,
to obtain the $M \times N$ matrix $\bD^q$ as 
\be \label{eqn:gen}
\bD^q = \bGamma_{\tau}^q \BD^q \bGamma_{\nu}^q,
\ee
with the elements $D^q[l,k]$ for $l=0,\ldots, M-1$ and $k=0,\ldots, N-1$.
The matrices $\bGamma_\tau^q$ and $\bGamma_\nu^q$ are formed by the columns of $\I_{M}$ with the indices in $\mathbb{U}_\tau^q$ and the rows of  $\I_{N}$ with the indices from the set $\mathbb{U}_\nu^q$, respectively.
The generalized resource allocation approach was originally proposed in \cite{Farhang2024}.
%%%%%

%%%%%%%%%%%
\begin{figure}[!t]
\centering{\includegraphics[scale=0.45]{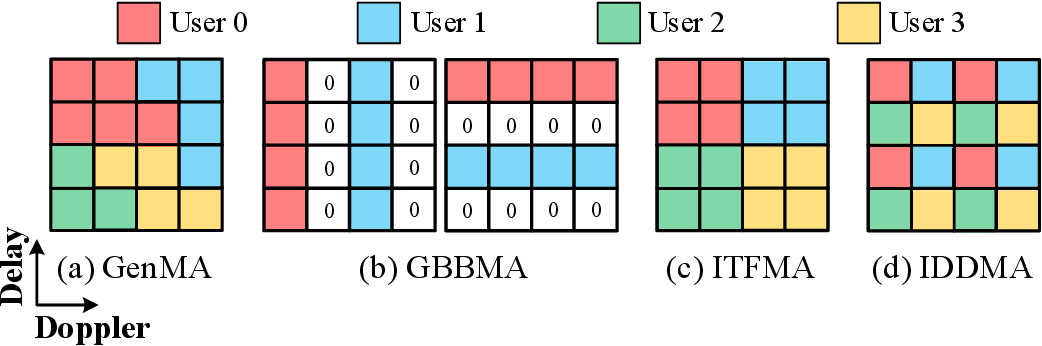}}
\vspace{-0.4cm}
\caption{Different multiple access schemes for OTFS.}
\vspace{-0.45cm}
\label{fig:MU_Schemes}
\end{figure}
%%%%%%%%%%%

This generalized multiple access (GenMA) scheme can accommodate all existing multiple access schemes for OTFS, including guard-band-based multiple access (GBBMA) \cite{hadani2018multi}, interleaved time-frequency multiple access (ITFMA) \cite{Augustine2019}, and interleaved delay-Doppler multiple access (IDDMA) \cite{Khan2019}, see Fig.~\ref{fig:MU_Schemes}.
In GBBMA, guard bands are inserted in the delay-Doppler domain to separate the information symbols transmitted by each MT. Two variants of GBBMA exist, including Doppler-domain GBBMA and delay-domain GBBMA. In Doppler-domain GBBMA, delay-Doppler resource elements assigned to different MTs are separated by guard bands along the Doppler domain. Conversely, in delay-domain GBBMA, the separation occurs along the delay dimension \cite{Kham2022, hadani2018multi}. The allocation of delay-Doppler resource elements for this scheme follows (\ref{eqn:gen}), with modifications to the resource allocation matrices. Specifically, for delay-domain GBBMA, the resource allocation matrices in (\ref{eqn:gen}) are defined as $\bGamma_{\tau}^q=\I_{M_q}$ and $\bGamma_{\nu}^q=\I_{N_q} \otimes [\bzero^{\rm{T}}_{q \times 1},1,\bzero^{\rm{T}}_{(Q-1-q) \times 1}]$, where $M_q=M$ and $N_q=\lfloor \frac{N}{Q} \rfloor$. For Doppler-domain GBBMA, the matrices are given by $\bGamma_{\tau}^q=\I_{M_q} \otimes [\bzero^{\rm{T}}_{q \times 1},1,\bzero^{\rm{T}}_{(Q-1-q) \times 1}]^{\rm{T}}$ and $\bGamma_{\nu}^q=\I_{N_q}$, where $M_q=\lfloor \frac{M}{Q} \rfloor$ and $N_q=N$.

\begin{figure}[!t]
\centering{\includegraphics[scale=0.45]{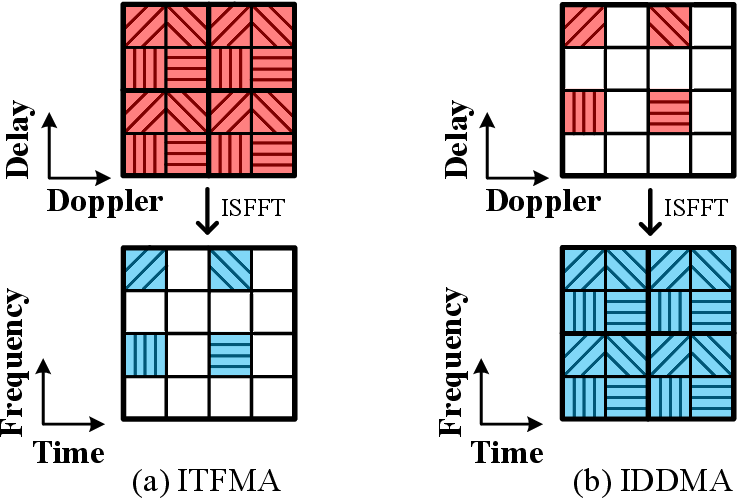}}
\caption{Data structure of a user at the MT for uplink transmission in both the delay-Doppler and time-frequency domains under a multiuser scenario.}
\vspace{-0.4cm}
\label{fig:MU_User1}
\end{figure}

In ITFMA, each user is allocated time-frequency resources in an interleaved manner. To achieve this interleaved frame structure in the time-frequency domain, each user repeats its modulated symbols across both the delay and Doppler dimensions before conversion to the time-frequency domain \cite{Augustine2019, Sinha2020, Das2024}. The transmitted frame structure of a user in the ITFMA uplink system for $M=4$ and $N=4$, which supports four simultaneous users, is shown in Fig.~\ref{fig:MU_User1}(a).
This method can also be considered a special case of (\ref{eqn:gen}). Specifically, the matrices $\bGamma_{\tau}^q = [\bzero_{M_q \times qM_q},\I_{M_q},\bzero_{M_q \times (\lfloor \frac{M}{M_q} \rfloor -1-q)M_q}]^{\rm{T}}$ and  $\bGamma_{\nu}^q=[\bzero_{N_q \times qN_q},\I_{N_q},\bzero_{N_q \times (\lfloor \frac{N}{N_q} \rfloor-1-q)N_q}]$, allocate resources along the delay and Doppler dimensions, respectively, for user $q$.
% 
%%%%%%%%%%%
%%%%%%%%%%%
% 
In IDDMA, each user is allocated delay-Doppler resources in an interleaved manner. This interleaved structure in the delay-Doppler domain results in each user repeating its modulated symbols across both the time and frequency dimensions in the time-frequency domain \cite{Kham2022, Khan2019}. The transmitted frame structure of a user in the IDDMA uplink system for $M=4$ and $N=4$, which supports four simultaneous users, is illustrated in Fig.~\ref{fig:MU_User1}(b).
This method can also be considered a special case of the resource allocation matrices in (\ref{eqn:gen}). Specifically, if the resource allocation matrices in (\ref{eqn:gen}) simplify to $\bGamma_{\tau}^q=\I_{M_q} \otimes [\bzero^{\rm{T}}_{q \times 1},1,\bzero^{\rm{T}}_{(\lfloor \frac{M}{M_q} \rfloor-1-q) \times 1}]^{\rm{T}}$ and $\bGamma_{\nu}^q=\I_{N_q} \otimes [\bzero^{\rm{T}}_{q \times 1},1,\bzero^{\rm{T}}_{(\lfloor \frac{N}{N_q} \rfloor -1-q) \times 1}]$, for the given user $q$.
%%%%%

Using the generalized resource allocation in (\ref{eqn:gen}), the transmit signal of user $q$ is formed by converting the data symbols into the delay-time domain via an inverse DFT (IDFT) operation along the Doppler dimension, i.e., as
\begin{align} \label{eqn:ifft} \bX^q = \bD^q  \bF_N^{\rm{H}}, \end{align}
with the elements of $\bX^q$ represented as
$X^q[l,n] = \frac{1}{\sqrt{N}} \sum_{k=0}^{N-1} D^q[l,k] e^{j \frac{2 \pi kn}{N}}$ for time index $n=0,\ldots,N-1$.
After parallel-to-serial conversion, $\x^q = {\rm{vec}} \{ \bX^q \}$, a CP of length $L_{\rm{cp}}$ is appended at the beginning of the OTFS frame for user $q$ as
\begin{align} \label{eqn:Acp} \s^q = \bA_{\rm{cp}}  \x^q, \end{align}
where $\bA_{\rm{cp}}=[\bJ_{\rm{cp}}^{\rm{T}},\bI^{\rm{T}}_{MN}]^{\rm{T}}$ is the CP addition matrix and the matrix $\bJ_{\rm{cp}}$ includes the last $L_{\rm{cp}}$ rows of the identity matrix $\bI_{MN}$. We consider a quasi-synchronous system in time where the signals from MTs arriving at the BS are time-aligned within the CP \cite{Morelli2007}.
In a quasi-synchronous system, the channel length for user $q$ extends from $L^q_{\rm{ch}}$ to $\check{L}^q_{\rm{ch}}\!=\!L^q_{\rm{ch}}+\theta^q$, where $\theta^q$ is the TO of user $q$, normalized by the delay spacing.
Thus, the CP length can be chosen as $L_{\rm{cp}} \!=\! \max_q \{ \check{L}^q_{\rm{ch}} \} \!=\! \max_q \{ L^q_{\rm{ch}} \} + \theta_{\rm{max}} - 1$, where $\theta_{\rm{max}} = \frac{\max_q \{ \Delta t^q \}}{\Delta \tau}$. Additionally, $\Delta t^q = \tau_{\rm{max}} + \Delta t^q_0$, where $\tau_{\rm{max}}$ denotes the maximum delay spread of the channel, and $\Delta t^q_0$ accounts for timing drift caused by the relative mobility of the transmitter and receiver for user $q$.
Substituting (\ref{eqn:ifft}) into (\ref{eqn:Acp}), the transmit signal can be obtained as $\s^q \!=\! \bA_{\rm{cp}} (\bF_N^{\rm{H}} \!\otimes\! \bI_M) \bd^q$
where $\bd^q \!\!=\! \bGamma^q \check{\bd}^q$, $\check{\bd}^q \!\!=\! {\rm vec}\{\BD^q\}$ and $\bGamma^q \!=\! (\bGamma^{q}_{\nu})^{\rm{T}} \otimes \bGamma_\tau^q$.

Considering multiple TO and CFO effects of the received signal from all MTs at the BS can be represented as
\begin{align} \label{eqn:rec20}
r[\kappa] &= \sum_{q=0}^{Q-1} e^{j \frac{2 \pi \varepsilon^q}{N_{\rm{s}}}\kappa} \sum_{\ell=0}^{L^q_{\rm{ch}}-1} s^q[\kappa-\ell-\theta^q] h^q[\ell,\kappa] + \eta[\kappa],
\end{align}
where $\kappa=0,\ldots,N_{\rm{s}}-1$ and $N_{\rm{s}}=MN+L_{\rm{cp}}$ is the total number of samples for an OTFS frame over a duration of $T_{\rm{f}}$ seconds.
The parameter $\varepsilon^q$ represents the CFO of user $q$, normalized by the Doppler spacing.
In (\ref{eqn:rec20}), $h^q[\ell,\kappa]$ is the impulse response of the LTV channel between the user $q$ and BS antenna, expressed as
\be \label{eqn:ch0} h^q[\ell,\kappa]=\sum_{i=0}^{\Upsilon^q-1} h^q_i e^{j 2 \pi \nu_i^q (\kappa-\ell)} \delta[\ell-\ell^q_i], \ee
where $h^q_i$, $\ell^q_i$, and $\nu^q_i=\nu_{\rm{max}} \cos \psi^q_i$, are channel gains, delays, maximum Doppler shifts, and the angle of arrival, respectively, for path $i$ and user $q$.
The maximum Doppler shift normalized to the Doppler spacing is also defined as $\kappa_{\max} = \nu_{\rm{max}} T_{\rm{f}} = \nu_{\rm{max}} N T = \nu_{\rm{max}} N M \Delta \tau$.
Additionally, $\Upsilon^q$ is the total number of paths and $\eta[\kappa] \!\sim\! \mathcal{CN} (0,\sigma^2_\eta)$ is the complex additive white Gaussian noise (AWGN) with the variance $\sigma^2_\eta$.
By substituting (\ref{eqn:ch0}) into (\ref{eqn:rec20}), the received signal can be rearranged as
\begin{align} \label{eqn:rec2}
r[\kappa]
&= \sum_{q=0}^{Q-1} \sum_{\ell=0}^{L^q_{\rm{ch}}-1} \check{h}^q[\ell,\kappa] s^q[\kappa-\ell-\theta^q] + \eta[\kappa],
\end{align}
where $\check{h}^q[\ell,\kappa]=\sum_{i=0}^{\Upsilon^q-1} \check{h}^q_i e^{j 2 \pi \check{\nu}_i^q (\kappa-\ell)} \delta[\ell-\ell^q_i]$ with $\check{h}^q_i=h^q_i e^{j 2 \pi \delta^q \ell}$, $\check{\nu}^q_i=\delta^q+\nu_i^q$, and $\delta^q=\frac{\varepsilon^q}{N_{\rm{s}}}$.

Equation (\ref{eqn:rec2}) shows that the CFO of each user is added as a constant to all the Doppler shifts of its channel, increasing the maximum Doppler frequency of the channel \cite{Das2021}. 
This effect makes channel estimation more challenging, especially as $|\varepsilon^q|$ increases.
From (\ref{eqn:rec2}), one may realise that the CFO and TO for each user can be absorbed into its channel and thus, can be estimated as a part of the channel and compensated at the equalization stage.
The TOs though need to be compensated so that we can find the location of the pilot for each user on the delay-Doppler plane.
As shown in Section~\ref{sec:Result}, if the CFOs are estimated as a part of the channel, the channel estimates are not as accurate as estimating and compensating the CFOs first and then estimating the channel.
%%%%%%%%%
From (\ref{eqn:rec2}), it is evident that the CFO of each user appears as an additive shift to all Doppler taps of the channel, effectively increasing the maximum Doppler spread. As a result, the total number of Doppler bins required to represent the channel increases, which degrades the sparsity of the delay-Doppler channel. If CFO is not estimated and compensated separately, the channel estimator must handle a wider Doppler spread, requiring a larger number of parameters to be estimated. This not only increases the estimation complexity but also reduces accuracy due to leakage across Doppler bins caused by multiuser interference. Moreover, the CFO-induced phase rotation varies over time, resulting in a time-varying delay-Doppler channel $\check{h}^q[\ell, \kappa]$, which violates the quasi-stationary assumption typically made in OTFS systems. In contrast, separating CFO estimation and compensation per user before channel estimation restores the sparsity and quasi-stationary of the channel in the delay-Doppler domain. While it is true that this approach incurs additional computational cost due to CFO estimation, this cost can be controlled by designing low-complexity CFO estimators, as demonstrated in this work.
%%%%%%%%%
Hence, in the following sections, we propose a CFO estimation technique for the MU-OTFS uplink.
Before proposing the techniques for estimating TO and CFO values, let us first investigate the impact of these offsets on the equivalent channel in the delay-Doppler domain. To understand the real effects of TO and CFO on the MU-OTFS channel, we derive the compound channel model for an asynchronous MU-OTFS system in the following section.

\vspace{-0.15cm}
\section{Channel Effect for MU-OTFS in Uplink}\label{sec:Ch}
\vspace{-0.05cm}
%%%%%%%%%%%%%%%%%%%
Let us consider the received signal at the BS from all MTs after transmission over the LTV channel by stacking the values of $r[\kappa]$, $s^q[\kappa]$, and $\eta[\kappa]$ for $\kappa=\theta_{\rm{max}},\ldots,N_{\rm{s}}-1$ into the vectors $\br$, $\s^q$, and $\boldsymbol{\eta}$, and using (\ref{eqn:rec20}) and (\ref{eqn:rec2}), the received signal vector can be represented as
\begin{align} \label{eqn:rec} \br &= \sum_{q=0}^{Q-1} \bPhi(\varepsilon^q) \bPi(\theta^q) \bH^q \s^q + \boldsymbol{\eta} \nonumber \\ &= \sum_{q=0}^{Q-1} \bPhi(\varepsilon^q) \bPi(\theta^q)\bH^q \bA_{\rm{cp}} (\bF_N^{\rm{H}} \otimes \bI_M) \bd^q + \boldsymbol{\eta}, 
\end{align}
where
$\bPhi(\varepsilon^q)={\rm{diag}} [ e^{j \frac{2\pi \varepsilon^q (\theta_{\rm{max}})}{N_{\rm{s}}}},\ldots,e^{j \frac{2\pi \varepsilon^q (MN+L_{\rm{cp}}-1)}{N_{\rm{s}}}} ]$ and
$\bPi(\theta^q)=[\mathbf{0}_{(MN+\check{L}_{\rm{cp}}) \times \theta^q},\bI_{(MN+\check{L}_{\rm{cp}})},\mathbf{0}_{(MN+\check{L}_{\rm{cp}}) \times (\theta_{\rm{max}}-\theta^q)}]$
are the CFO and TO matrices, respectively. 
The $\bH^q \in \mathbb{C}^{N_{\rm{s}}}$ is a Toeplitz matrix that stacks the channel coefficients from (\ref{eqn:ch0}).
After removing the CP of length $\check{L}_{\rm{cp}}=L_{\rm{cp}}-\theta_{\rm{max}}$, the received signal can be expressed as $\y=\bR_{\rm{cp}} \br$ where  the CP removal matrix is defined as $\bR_{\rm{cp}}=[\mathbf{0}_{MN\times \check{L}_{\rm{cp}}},\bI_{MN}]$.
Hence, the received delay-Doppler domain signal is obtained by applying $N$-point DFT to $\y$ along the time dimension as 
\begin{align} \label{eqn:d_hat} \widetilde{\bd} \!&=\! (\bF_N \!\otimes\! \bI_M) \y = (\bF_N \otimes \bI_M) \bR_{\rm{cp}} \br \nonumber \\
&= \sum_{q=0}^{Q-1} (\bF_N \otimes \bI_M) \check{\bPhi}(\varepsilon^q) {\bLambda}^q \x^q + \check{\boldsymbol{\eta}} \nonumber \\
\!&=\!\!\! \sum_{q=0}^{Q-1} (\bF_N \otimes \bI_M) \check{\bPhi}(\varepsilon^q) {\bLambda}^q (\bF_N^{\rm{H}} \otimes \bI_M) \bd^q \!+\! \check{\boldsymbol{\eta}} \nonumber \\
&= \sum_{q=0}^{Q-1} \check{\bPhi}_{\rm{DD}}(\varepsilon^q) \bLambda^q_{\rm{DD}} \bd^q + \check{\boldsymbol{\eta}} = \bPsi_{\rm{DD}} \bd + \check{\boldsymbol{\eta}}.\end{align}
In (\ref{eqn:d_hat}), the CFO matrix of each user after CP removal can be presented by $\check{\bPhi}(\varepsilon^q)\!=\!{\rm{diag}}[e^{j \frac{2\pi \varepsilon^q (L_{\rm{cp}})}{N_{\rm{s}}}},\ldots,e^{j \frac{2\pi \varepsilon^q (MN+L_{\rm{cp}}-1)}{N_{\rm{s}}}}]$, ${\bLambda}^q \!=\! \bR_{\rm{cp}} \bPi(\theta^q) \bH^q \bA_{\rm{cp}}$ is the combination of TO and channel response for user $q$, and $\check{\boldsymbol{\eta}}=(\bF_N \otimes \bI_M) \bR_{\rm{cp}} \boldsymbol{\eta}$ is AWGN in the delay-Doppler domain.
In (\ref{eqn:d_hat}), $\check{\bPhi}_{\rm{DD}}(\varepsilon^q) = (\bF_N \otimes \bI_M) \check{\bPhi}(\varepsilon^q) (\bF^{\rm{H}}_N \otimes \bI_M)$ and $\bLambda^q_{\rm{DD}}=(\bF_N \otimes \bI_M) {\bLambda}^q (\bF_N^{\rm{H}} \otimes \bI_M)$ are the CFO matrix and the channel response of the $q^{\text{th}}$ user in the delay-Doppler domain, respectively.
Ultimately, $\bPsi_{\rm{DD}}\!=\!\sum_{q=0}^{Q-1} \! \check{\bPhi}_{\rm{DD}}(\varepsilon^q) \bLambda^q_{\rm{DD}} (\bGamma^q)^{\rm{H}}$ is the compound channel that includes the TO, CFO, and channel responses of all the users and $\bd \!=\! \sum_{q=0}^{Q-1} \bd^q \!=\! \sum_{q=0}^{Q-1} \bGamma^q \check{\bd}^q$ is the combined delay-Doppler domain transmitted data symbols of all the users.

From (\ref{eqn:d_hat}), one might deduce that the best way to compensate the effects of multiple TOs and CFOs in the uplink and equalize the channel is by using the compound channel matrix $\bPsi_{\rm{DD}}$. However, as mentioned, channel estimation after synchronization is more accurate than directly estimating the channel which includes the CFO and TO effects. This is because TO estimation is required to find the users' pilot signals and CFO compensation before channel estimation prevents unwanted channel estimation errors. To address the above challenges, the next section will propose two TO synchronization techniques for different pilot structures in MU-OTFS, followed by a joint CFO and channel estimation technique in Section~\ref{sec:ML}.

\vspace{-0.15cm}
\section{Proposed Correlation-Based TO Estimation Techniques} \label{sec:TO}
\begin{figure}[!t]
  \centering 
  {\includegraphics[scale=0.19]{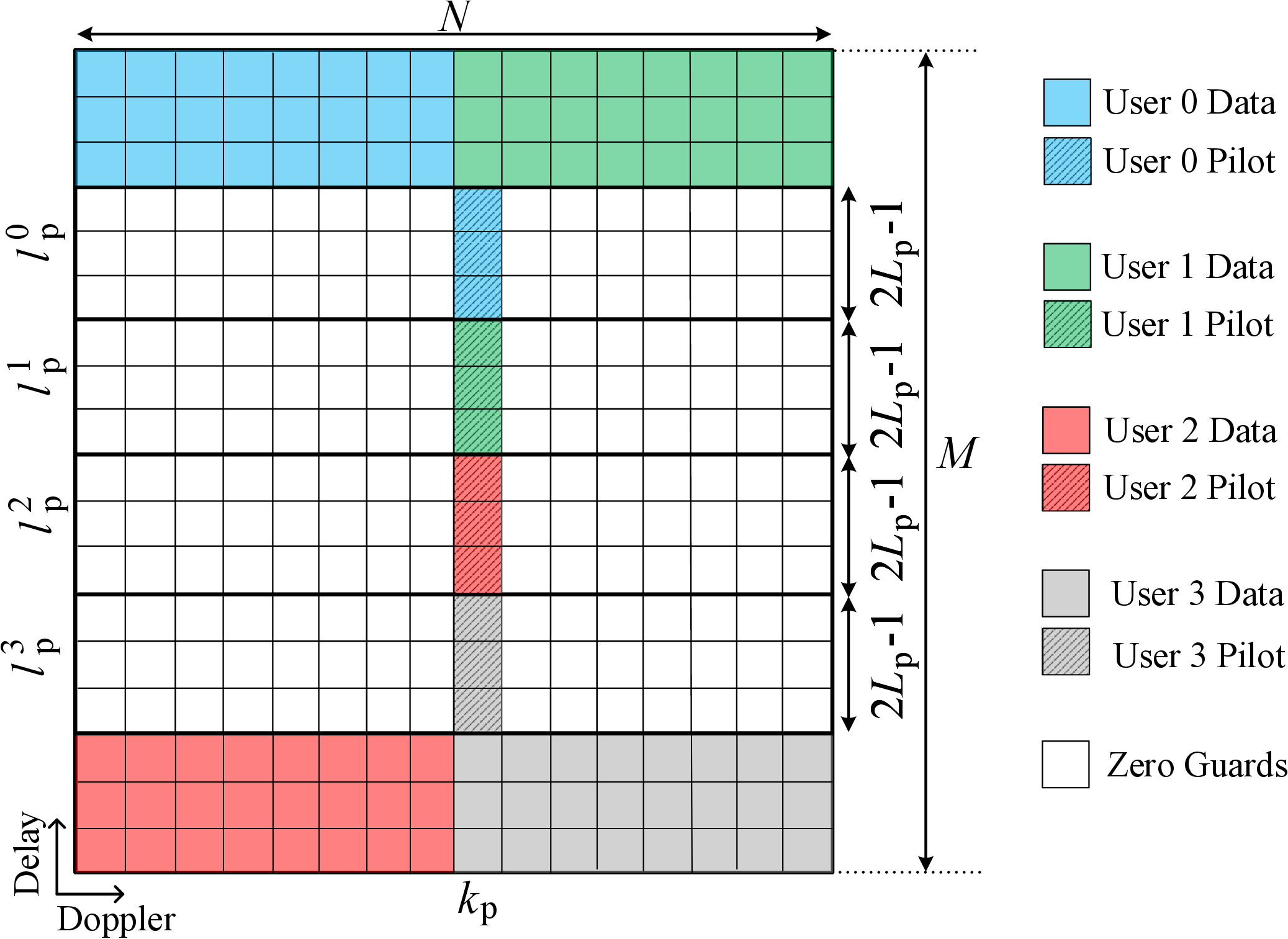}}
    \vspace{-0.2cm}
  \caption{SU-PCP structure for MU-OTFS in the delay-Doppler domain.}
  \vspace{-0.4cm}
  \label{fig:pilot0}
\end{figure}

\vspace{-0.05cm}
Due to timing errors in uplink transmission, an additional TO estimation stage is required in an uplink MU-OTFS system to locate the pilot region, which is used for CFO and channel estimation \cite{Bayat2022}. Hence, this section focuses on the problem of estimating multiple TOs in uplink MU-OTFS.
In Section~\ref{sec:TO1}, a TO estimation technique is developed for an existing multiuser pilot structure proposed in \cite{Raviteja2018}, using impulse pilot (IMP). However, the IMP in this structure is replaced with PCP from \cite{Sanoop2023} as a more practical solution, which leads to a modified structure, SU-PCP as shown in Fig.~\ref{fig:pilot0}.
In Section~\ref{sec:TO2}, MU-PCP a new pilot structure, is introduced, which utilizes a shared pilot region for all users to enhance spectral efficiency. Subsequently, a TO estimation technique is proposed for this new pilot structure.
The proposed technique uses a threshold to detect the first major peak of the correlation function. This threshold is also derived in Section~\ref{sec:TO_err}.

%%%%%%%
In general, the CP in CP-OTFS enables correlation-based TO estimation by introducing similarity between the beginning and end of each delay block in static channels \cite{hadani2017}. However, in high-mobility environments, channel time variations can distort this repetitive structure, reducing its effectiveness. This issue is further exacerbated in reduced-CP configurations that are commonly adopted to enhance spectral efficiency \cite{Raviteja2018}. To avoid the aforementioned issues, a short OFDM-like pilot sequence can be embedded within each OTFS delay block in the delay-time domain, such that the CP and the end of the OFDM symbol are close together. Thus, CP redundancy in PCP, which is used to reduce data interference on the pilot, can also be used to improve TO estimation accuracy. This motivates the use of a CP in PCP for synchronization, as it preserves the similarity between the repetitive signal parts, thereby improving TO estimation accuracy.
%%%%%%%
%%%%%%%%%%%%
%%%%%%%%%%%%

%%%%%%%%%%%%%%%%%%
\vspace{-0.1cm}
\subsection{TO Estimation Technique for SU-PCP} \label{sec:TO1}
\begin{figure}[!t]
% \vspace{-0.4cm}
  \centering 
  {\includegraphics[scale=0.28]{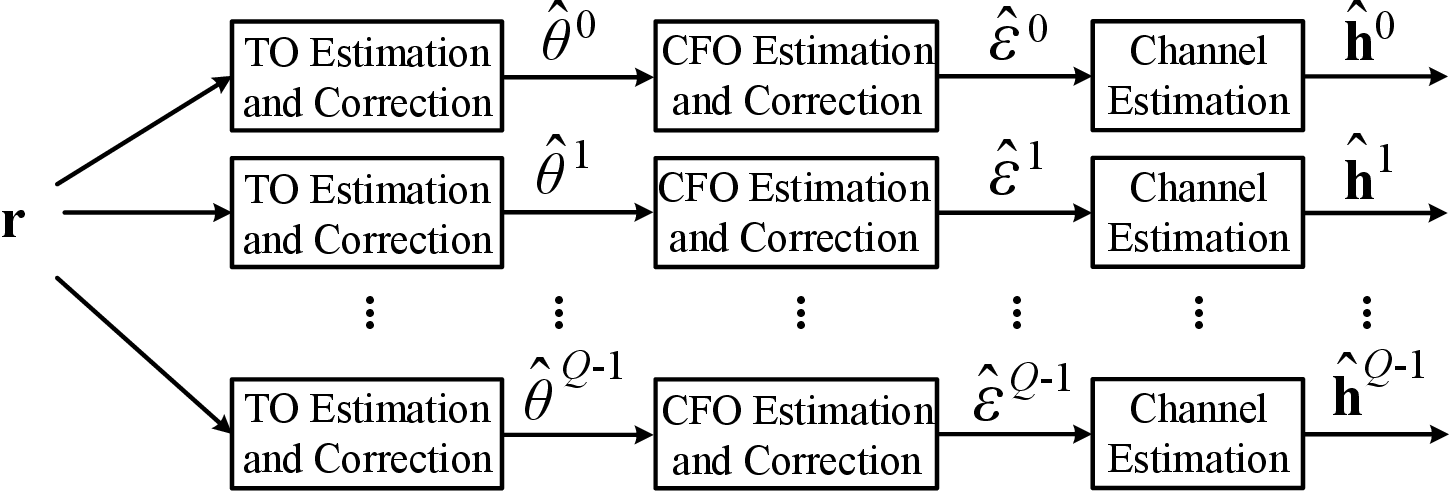}}
  \vspace{-0.3cm}
  \caption{The proposed synchronization block diagram for SU-PCP.}
  \vspace{-0.5cm}
  \label{fig:ML_MU20}
\end{figure}

\vspace{-0.05cm}
As shown in Fig.~\ref{fig:ML_MU20}, the entire problem can be formulated to utilize a correlation-based TO estimation to find TOs $\hat{\boldsymbol{\theta}}=[\hat{\theta}^0,\ldots,\hat{\theta}^{Q-1}]^{\rm{T}}$, using the designated pilot region for each user. Then, an ML estimator jointly estimates CFOs and channel coefficients, denoted by $\hat{\boldsymbol{\varepsilon}}=[\hat{\varepsilon}^0,\ldots,\hat{\varepsilon}^{Q-1}]^{\rm{T}}$ and $\hat{\bh}=[(\hat{\bh}^0)^{\rm{T}},\ldots,(\hat{\bh}^{Q-1})^{\rm{T}}]^{\rm{T}}$, respectively.

The  SU-PCP arranges the pilot regions of users consecutively along the delay dimension, while the remaining delay-Doppler bins are arbitrarily allocated for data transmission.
For each user, a PCP of length $2{L}_{\rm{p}}-1$ is generated using a predefined sequence, such as a ZC of length ${L}_{\rm{p}}$, where the last ${L}_{\rm{p}}-1$ elements are copied to the beginning as a CP. In the proposed pilot structure, the pilots for users $q=0,\ldots,Q-1$ are placed consecutively along the delay dimension, i.e., from $l^q_{\rm{p}}-{L}_{\rm{p}}+1,\ldots,l^q_{\rm{p}}+{L}_{\rm{p}}-1$, where $l^q_{\rm{p}} = l_{\rm{p}} + L_{\rm{p}} + q(2 {L}_{\rm{p}}-1)$ and $l_{\rm{p}}=\frac{M}{2} + \lfloor \frac{Q}{2} \rfloor - Q {L}_{\rm{p}}$, at the Doppler bin $k_{\rm{p}}=\frac{N}{2}$, see Fig.~\ref{fig:pilot0}.
As will be discussed shortly, to enable differentiation of the correlation functions of the received signals for different users' pilot signals, this pilot structure allocates different ZC sequences to different users.

In this multiuser pilot structure, the number of distinct ZC sequences that can be assigned to different users is limited by the sequence length, as they are generated using different root indices. Specifically, for a ZC sequence of length $L_{\rm{p}}$, up to $L_{\rm{p}}-1$ valid root indices exist, each yielding a distinct sequence. In addition, the pilot structure in SU-PCP allocates each user’s pilot across $2L_{\rm{p}}-1$ delay bins. Since the total number of delay bins in an OTFS frame is $M$, the number of users is further constrained by $\left\lfloor \frac{M}{2L_{\rm{p}}-1} \right\rfloor$. Therefore, the maximum number of distinguishable users in SU-PCP is the minimum of $(L_{\rm{p}}-1)$ and $\left\lfloor \frac{M}{2L_{\rm{p}}-1} \right\rfloor$. It is worth noting that, in this scenario, $L_{\rm{p}}=L_{\rm{ch}}$.

%%%%%%%%%%%%
%%%%%%%%%%%%

As mentioned earlier, multiple TO estimations in the uplink of MU-OTFS are necessary for locating users' pilot signals to perform CFO and channel estimation.
As the TOs are not estimated as part of the channel in this approach, the parameter $L_{\rm{p}}$ depends only on $\max_q\{L_{\rm{ch}}^q\}$, rather than $\max_q\{\check{L}_{\rm{ch}}^q\}$. This is a key advantage of the proposed method, as it allows for a shorter pilot compared to cases where TO is absorbed into the channel, thereby enhancing spectral efficiency.
If the TO is absorbed into the channel, the first non-zero tap of the channel does not appear at the first tap of the channel response. Instead, it appears at the tap $\theta_{\rm max}^q$ for a given user $q$. This will lead to an increased channel length from $L_{\rm ch}^q$ to $L_{\rm ch}^q+\theta_{\rm max}^q$ and we need to estimate a larger number of unknown elements in channel estimation stage. Consequently, the pilot length needs to be increased. Additionally, since the pilots of all the users need to be quasi-synchronous, the CP length for the pilot also needs to increase to the same length as for the OTFS frame. In contrast, if we first perform TO estimation and compensation, the CP length of the pilot needs to be $\max_q \{ L^q_{\rm{ch}} \}$.
Since the pilot structure presented in Fig.~\ref{fig:pilot0} inserts each user's pilot regions consecutively across the delay dimension, it requires only a pilot region of length $2L_{\rm{p}}\!-\!1 \!=\! 2\max_q \{ L^q_{\rm{ch}} \}\!-\!1$ for each user, instead of $2 \max_q \{ \check{L}^q_{\rm{ch}} \} - 1 = 2 (\max_q \{ L_{\rm{ch}}^q \} + \theta_{\rm{max}}) - 1$.

To find the periodicity within each user's pilot signal, it is required to slide a pure pilot signal including the users' pilot in its initial position to find the location of the received pilot for that user. This process is required to be repeated $Q$ times to estimate $\theta^q$ for $q=0,\ldots,Q-1$.
The transmitted pilot of each user in the delay-time dimension can be correlated and slid over the signal $\br$ to estimate $\theta^q$.
The first $MN$ samples of the received vector $\br$ can be reshaped into a matrix $\bR[l,n] = r[\kappa] = r[nM + l] \in \mathbb{C}^{M \times N}$, where $r[\kappa]$ is obtained in (\ref{eqn:rec2}), and $l = 0, \ldots, M - 1$ and $n = 0, \ldots, N - 1$.
Additionally, matrix $\bZ^q_{l'}[l,n] \in \mathbb{C}^{M \times N}$ represents the transmitted delay-time pilot signal
$\bZ^q=[\bzero_{(l^q_{\rm{p}}-L_{\rm{p}}+1) \times N},({\bO}^q \bF_N^{\rm{H}}),\bzero_{(M-l^q_{\rm{p}}-L_{\rm{p}}) \times N}]^{\rm{T}}$,
circularly shifted upwards by $l'$ positions along the delay dimension.
Given ${\bO}^q=[\bzero_{(2L_{\rm{p}}-1) \times \frac{N}{2}},\z^q_{\rm{p}},\bzero_{ (2L_{\rm{p}}-1) \times (\frac{N}{2}-1)}]$ and $\z^q_{\rm{p}}=[(\z^q)^{\rm{T}},(\z^q)^{\rm{T}}]^{\rm{T}}$ is a vector of size $(2L_{\rm{p}}-1) \times 1$, containing the PCP sequence elements in the delay-Doppler plane.
In the case of a ZC sequence \cite{3gpp_5}, $\z^q$ stacks the elements of $z_q[\iota] = \sqrt{N \sigma^2_{\rm{p}}} e^{-j \frac{\pi \mu \iota (\iota+1)}{L_{\rm{p}}} (2q+1)}$ for $\iota=0,\ldots,L_{\rm{P}}-1$ and $q=0,\ldots,Q-1$, where $\sqrt{N \sigma^2_{\rm{p}}}$ represents the pilot power, and $\mu$ is the root index satisfying  ${\rm{gcd}}(L_{\rm{p}},\mu)=1$.
Then, the timing metric along the delay dimension for TO estimation can be formulated as
\be \label{eqn:TO1_Corr} \p^q [l'] = \frac{1}{MN} \sum_{n=0}^{N-1} \Big| \sum_{l=0}^{M-1} \bR[l,n] \odot (\bZ^q_{l'}[l,n])^{\rm{H}} \Big|, \ee
where $l'=0,\ldots,M-1$.

As explained in \cite{Bayat2023} and \cite{Farhang2024}, the peak of the correlation function for TO estimation is primarily influenced by the strongest tap of the channel.
Hence, identifying the first major peak of the correlation function provides a TO estimate for each user. 
Accordingly, we propose to find the first major peak of $\p^q[l']$.
In this paper, we use the cross-correlation function of the received signal and the pilot signal as described in (\ref{eqn:TO1_Corr}).
To find the first major peak, we group all the peaks using a threshold value $\mathcal{T}^q$, such that $0<\mathcal{T}^q < 1$.
The constructed set of peaks can be described mathematically as
\be \label{eqn:TO_d} 
\boldsymbol{\widehat{\Theta}}^q \!=\! \Big\{ l' \, \Big| \, \big|\p^q[l'] \big| \!\geq\! \big( \mathcal{T}^q \times \max \big\{ |\p^q[l']| \big\} \big) \Big\}.
\ee
Consequently, the first major peak is identified as
\be \label{eqn:TO_d2} 
\hat{\theta}^q= \big( \min \{ \boldsymbol{\widehat{\Theta}}^q \}-L_{\rm{cp}}-l^q_{\rm{p}} +1 \big)_M.
\ee
The correlation of a sliding PCP over the received PCP in each delay block generally produces three major peaks due to the structure of the PCP, in addition to the minor peaks caused by the ZC sequence structure. Two of the major peaks in the correlation function result from the alignment of the CP in the PCP, while the third arises from full alignment of the entire PCP sequence. The objective of the proposed TO estimation technique is to identify the first major peak in the correlation function along the delay dimension, see Fig.~\ref{fig:Mean}-(c). This can be achieved by selecting an appropriate detection threshold.
The derivation of the threshold $\mathcal{T}^q$ for determining an appropriate value will be presented in Section~\ref{sec:TO_err}.

Alternatively, rather than detecting the first peak, the highest peak can be used for TO estimation as
\be \label{eqn:highest_peak} 
\hat{\theta}^q = \big(\arg\max_{l'} \big\{ |\p^q[l']| \big\} - L_{\rm{cp}} - l^q_{\rm{p}} - L_{\rm{p}} - \lambda^q_{\rm{ch}} +1 \big)_M,
\ee
where $\lambda^q_{\rm{ch}} \!\!=\!\! \big \lfloor \! \sum_{i=0}^{L^q_{\rm{ch}}\!-\!1} \!\!\ell^q_i |h^q_i|^2 \!\big/\! \sum_{i=0}^{L^q_{\rm{ch}}\!-\!1} \! |h^q_i|^2 \! \big \rfloor$ represents the channel mean delay, normalized by the delay spacing, i.e., $\ell_i = \frac{\tau_i}{\Delta \tau}$.

Typically, the amplitude of the highest peak using PCP in the correlation function is very close to that of its neighboring peaks, especially in the region where the sliding transmit pilot begins to overlap with the received pilot. The highest peak generally occurs when the transmit and received pilots are perfectly aligned. However, adjacent peaks, where the transmit pilot overlaps with almost all rows of the received pilot, can also produce nearly the same correlation values. In highly time-varying channels, this may result in attenuation of the actual peak and amplification of adjacent peaks, potentially leading to inaccurate TO estimation. In contrast, selecting the first major peak within the pilot region helps mitigate this issue. Due to the structure of the PCP formed from a ZC sequence, this peak occurs at a position in the correlation function that stands out sharply above the preceding values. As a result, the correlation values before this first major peak remain relatively low, even under severe channel conditions. This situation, makes the first major peak more distinguishable and reliable for TO estimation.
In channels where the strongest tap arrives later, the highest peak may still provide a valid estimate, as will be shown in Section~\ref{sec:Result}, however, its performance degrades compared to scenarios where the strongest tap is the first. In such cases, as shown in (\ref{eqn:highest_peak}), the TO estimate obtained from the highest peak can be improved by adjusting it using the channel delay mean, denoted by $\lambda^q_{\rm{ch}}$.
Additionally, Section~\ref{sec:Result} will demonstrate that, in these scenarios, using the first major peak for TO estimation yields more robust results than relying on the highest peak.

As observed in Fig.~\ref{fig:pilot0}, there is still potential to make the multiuser pilot structure more spectrally efficient. To this end, the next subsection presents a spectrally efficient pilot structure for MU-OTFS, which allocates a shared pilot region for all users in delay. This pilot structure requires an update to the TO estimation technique described in this subsection.

%%%%%%%%%%%%%%%%%%
\vspace{-0.15cm}
\subsection{TO Estimation Technique for MU-PCP} \label{sec:TO2}
\begin{figure}[!t]
  \centering 
  {\includegraphics[scale=0.19]{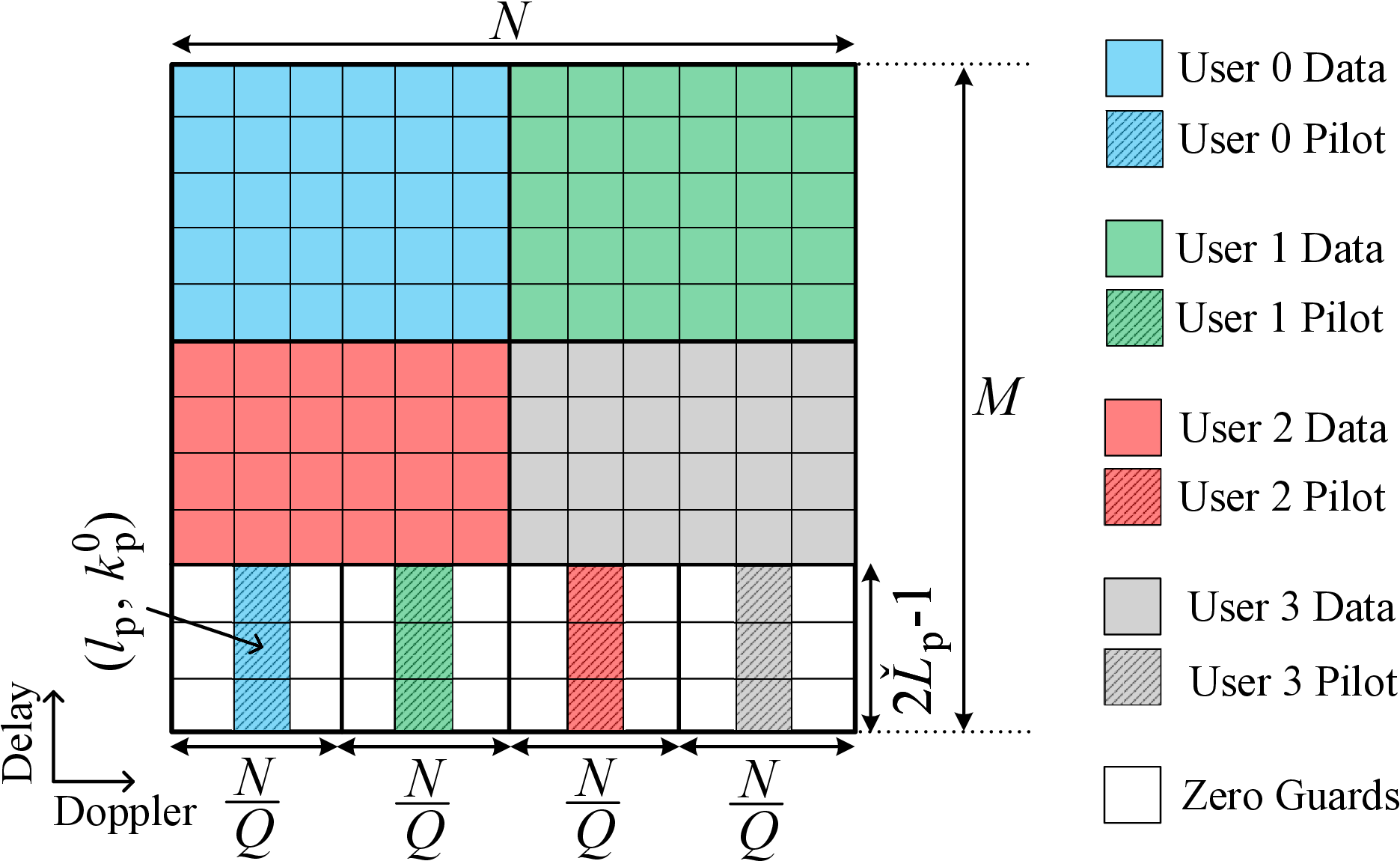}}
  \vspace{-0.35cm}
  \caption{Proposed MU-PCP structure for MU-OTFS in the delay-Doppler domain.}
  \vspace{-0.4cm}
  \label{fig:pilot}
\end{figure}

\vspace{-0.05cm}
To enhance spectral efficiency for the multiuser pilot structure used the synchronization technique proposed in Section~\ref{sec:TO1}, this section introduces an alternative TO synchronization technique built on the spectrally efficient multiuser pilot structure, called MU-PCP.
In this pilot structure, a designated pilot region is proposed where all the users transmit their pilots, and the remaining delay-Doppler bins are arbitrarily allocated to the users for data transmission, see Fig.~\ref{fig:pilot}.
To estimate these parameters, the PCP is deployed as a pilot for synchronization, as originally proposed for channel estimation in \cite{Sanoop2023}.
For each user, a PCP of length $2\check{L}_{\rm{p}}-1$ is formed by a pre-defined sequence, i.e., ZC, of length $\check{L}_{\rm{p}}$ in which the last $\check{L}_{\rm{p}}-1$ elements of the sequence are copied at the beginning as a CP.
In this proposed pilot structure, the pilots of the users $q=0,\ldots,Q-1$ reside within the same delay region, i.e., $l = l_{\rm{p}}-\check{L}_{\rm{p}}+1,\ldots, l_{\rm{p}}+\check{L}_{\rm{p}}-1$ in Doppler bin $k^q_{\rm{p}} = \frac{1}{2} \lfloor \frac{N}{Q} \rfloor + q \lfloor \frac{N}{Q} \rfloor$, see Fig.~\ref{fig:pilot}.
Unlike SU-PCP, MU-PCP does not require different ZC sequences for different users.

In the proposed MU-PCP structure, the number of users that can be accommodated along the Doppler dimension ranges from a minimum of 1 up to the total number of Doppler bins, i.e., $N$. A key factor limiting this number is the Doppler spread of the time-varying channel, which constrains the user capacity to $Q \leq \left\lfloor \frac{N}{4\kappa_{\max} + 1} \right\rfloor$, where $\kappa_{\max}=\nu_{\max}T_{\rm{f}}$. However, by incorporating a bandpass filter into the proposed framework, the system can more effectively manage worst-case Doppler conditions. In practice, most of the Doppler-domain channel energy is concentrated within a central portion of the total spread. For instance, under the modified Jakes model, such as raised cosine Doppler spectrum, more than $90\%$ of the Doppler-domain power is typically confined within $50\%$ of the total Doppler span, as shown in Appendix~A of the revised manuscript. Based on this, we define the dominant Doppler spread as $\kappa_{\rm{d}} = \alpha \kappa_{\max}$, where $\alpha \approx 0.5$, serving as a practical approximation of the effective Doppler support in system design. Hence, we could accommodate a larger number of users by relaxing the maximum Doppler constraint and considering only the dominant Doppler energy of each user. Accordingly, the updated upper bound on the number of users in the MU-PCP structure becomes $Q \leq \left\lfloor \frac{N}{4 \kappa_{\rm{d}} + 1} \right\rfloor$.

%%%%%%%%%%%%
%%%%%%%%%%%%

\begin{figure}[!t]
  \centering 
  {\includegraphics[scale=0.26]{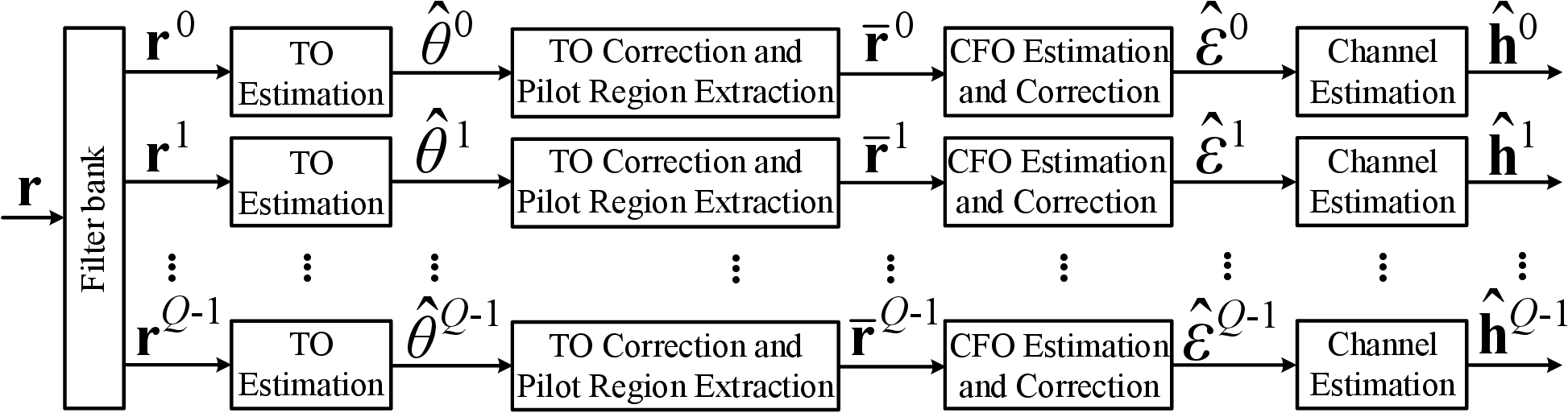}}
  \vspace{-0.1cm}
  \caption{The proposed structure for TO, CFO, and channel estimation using a bank of bandpass filters for MU-PCP.}
  \vspace{-0.2cm}
  \label{fig:FB}
\end{figure}

To estimate the TO of each user, we propose to pass the received signal $\br$ through a bank of digital bandpass filters that separates different users' pilots, see Fig.~\ref{fig:FB}.
Considering the brickwall filter $\ba^q = (\!(\ba)\!)_{q \lfloor \frac{N}{Q} \rfloor }$ where $\ba=[\bone^{\rm{T}}_{\lfloor \frac{N}{Q} \rfloor \times 1},\bzero^{\rm{T}}_{(Q-1)\lfloor \frac{N}{Q} \rfloor \times 1}]^{\rm{T}}$, the signal of a given user $q$ can be separated by multiplication of the received signal in Doppler by the filter response which is equivalent to circular convolution in delay-time domain by $\e^q = \bF^{\rm{H}}_N \ba^q$. Hence, circular convolution matrix $\bE^q={\rm{circ}} \{ \e^q \}$ can be used as follows to obtain $\br^q$.
This enables the BS to independently estimate the TOs for different users and thus, reduce a $Q$-dimensional search problem into $Q$ parallel one-dimensional search problems. 
The separated pilot signal for a given user $q$ can be expressed as
\be \label{eqn:BF} \br^q = \big( (\bE^q)^{\rm{H}} \otimes \bI_M \big) \check{\br}, \ee
where $q=0,\ldots,Q-1$ and $\check{\br}$ represents the first $MN$ elements of the received vector $\br$ in (\ref{eqn:rec}).

%%%%%%%%%%%%
%%%%%%%%%%%%

\begin{figure}[!t]
  \centering 
  {\includegraphics[scale=0.175]{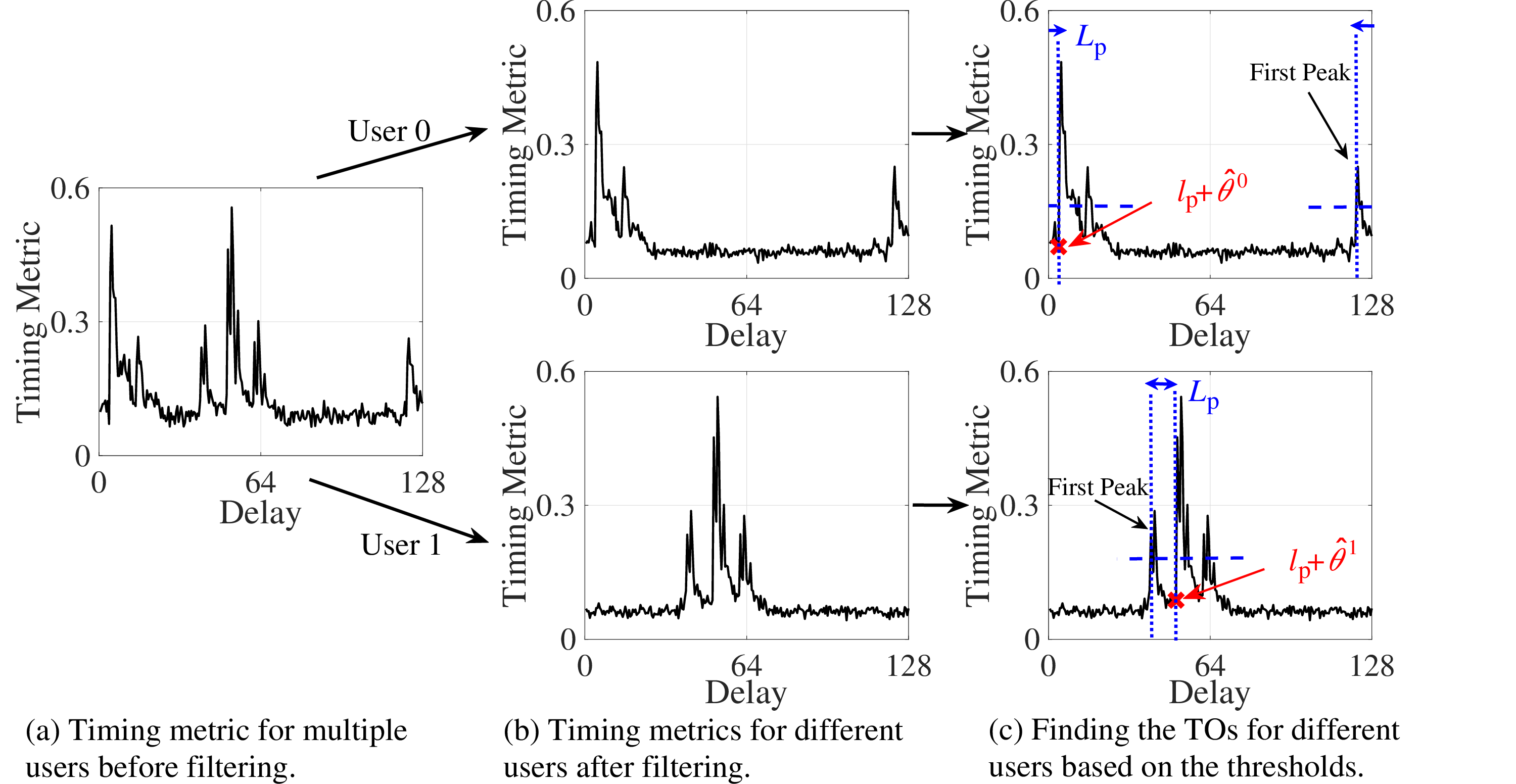}}
  \vspace{-0.7cm}
  \caption{A snapshot of the timing metrics for $Q=2$ and $\kappa_{\rm{max}} \approx 2.91 \,\, \forall q$.}
  \vspace{-0.4cm}
  \label{fig:Mean}
\end{figure}

After the separation of different pilot signals of the users, the TOs for different users can be estimated in parallel by finding the two identical halves of PCP in the delay-time domain. 
Hence, the transmitted pilot of each user in the delay-time dimension can be correlated and slid over the signal $\br^q$ to estimate $\theta^q$ for $q=0,\ldots,Q-1$, see Fig.~\ref{fig:Mean}-(a) and -(b).
The timing metric for TO estimation can be obtained as
\be \label{eqn:TO2_Corr} \check{\p}^q [l'] = \frac{1}{MN} \sum_{n=0}^{N-1} \Big| \sum_{l=0}^{M-1} \check{\bR}^q[l,n] \odot (\check{\bZ}^q_{l'}[l,n])^{\rm{H}} \Big|, \ee
where $l'=0,\ldots,M-1$ and $\check{\bm{R}}^q[l,n] = \bm{r}^q[nM+l] \in \mathbb{C}^{M \times N}$ is constructed using the received filtered vector $\bm{r}^q$ in (\ref{eqn:BF}) for $l=0,\ldots,M-1$ and $n=0,\ldots,N-1$. 
The matrix $\check{\bm{Z}}^q_{l'}[l,n] \in \mathbb{C}^{M \times N}$ represents the transmitted delay-time pilot signal $\check{\bm{Z}}^q$, circularly shifted upward by $l'$ positions along the delay dimension.
Here, $\check{\bm{Z}}^q$ is defined as $\check{\bm{Z}}^q=[\bm{0}_{(l_{\rm{p}}-\check{L}_{\rm{p}}+1) \times N},(\check{\bm{O}}^q \bm{F}_N^{\rm{H}}),\bm{0}_{(M-l_{\rm{p}}-\check{L}_{\rm{p}}) \times N}]^{\rm{T}}$, where $\check{\bm{O}}^q$ contains the pilot values in the delay-Doppler domain and multiplication by $\bm{F}_N^{\rm{H}}$ transforms them to the delay-time domain.
Given $\check{\bO}^q=[\bzero_{(2\check{L}_{\rm{p}}-1) \times (k^q_{\rm{p}}\!-1)},\check{\z}_{\rm{p}},\bzero_{ (2\check{L}_{\rm{p}}-1) \times (N-k^q_{\rm{p}})}]$ and $\check{\z}_{\rm{p}}=[\check{\z}^{\rm{T}},\check{\z}^{\rm{T}}]^{\rm{T}}$ is a vector of size $(2\check{L}_{\rm{p}}-1) \times 1$, containing the PCP sequence elements in the delay-Doppler plane.
In the case of a ZC sequence \cite{3gpp_5}, $\check{\z}$ stacks the elements of $z_0[\iota]$ for $\iota=0,\ldots,\check{L}_{\rm{P}}-1$.
Then, $\hat{\theta}^q$ can be obtained using (\ref{eqn:TO_d}), (\ref{eqn:TO_d2}), and (\ref{eqn:highest_peak}) by replacing $\p^q[l']$ with $\check{\p}^q[l']$, and assuming $l_{\rm{p}}=l^q_{\rm{p}}, \,\, \forall q$.

%%%%%%%%%%%%%%%%%%%%%%%%%%%%
\vspace{-0.05cm}
\subsection{Threshold Calculation for TO Estimation} \label{sec:TO_err}
To extract the major peaks of the correlation function caused by the embedded similarity in PCP, a threshold must be carefully chosen.
To choose the best values for the thresholds, it is necessary to find a valid range for these thresholds. Hence, in this subsection, we derive an equation for threshold selection in subsection~\ref{sec:TO2}.
Considering (\ref{eqn:rec20}) and given $\kappa=nM+l$ for $k=0,\ldots,M-1$ and $n=0,\ldots,N-1$, the timing metric in (\ref{eqn:TO2_Corr}) can be rewritten as
\begin{align} \label{eqn:corrq} 
\check{p}^q[l'] &= \frac{1}{MN} \sum_{n=0}^{N\!-\!1} \Big| \sum_{l=0}^{M\!-\!1} \check{R}^q[l,n] \check{Z}^{q^*}[(l\!-\!l')_M,n] \Big|,
\end{align}
%%%%%%%%%%%%
%%%%%%%%%%%%
where $\check{R}^q[l,n]=r^q[nM+l]=\sum_{\ell=0}^{L_{\rm{ch}}-1} h^q[\ell,nM+l] S^q[l-\theta-\ell,n]+\zeta[l,n]$, $S^q[l,n]=s^q[nM+l]$ stacks the elements of $\s^q$, and $\zeta[l,n]=\zeta[nM+l]=\eta^q[nM+l] \!\sim\! \mathcal{CN} (0,\sigma^2_\zeta)$ is an AWGN with variance $\sigma^2_{\zeta}=\frac{\sigma^2_{\eta}}{Q}$.
Since the received signal has passed through a bandpass filter, we can disregard interference from the pilots of other users.
Thus, for ease of presentation, we consider the first user $q=0$, and omit the superscript of $q$ in the remainder of this subsection.

%%%%%%%%%%%%
\begin{figure*}[!t]
% \vspace{-0.2cm}
  \centering 
  {\includegraphics[scale=0.19]{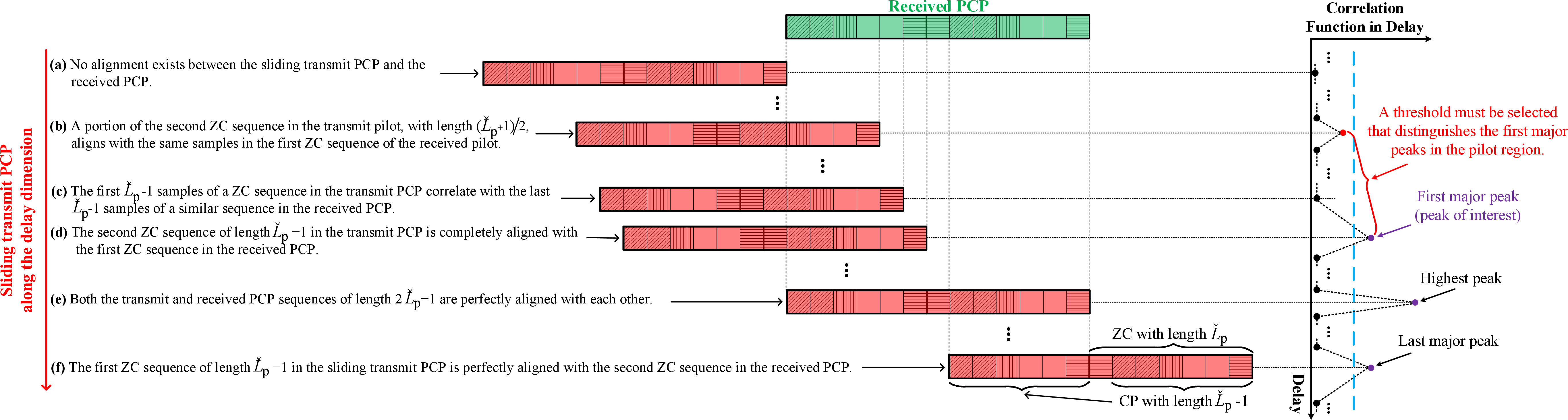}}
  \vspace{-0.5cm}
  \caption{Three major peaks in the pilot region relating to PCP.}
  \vspace{-0.5cm}
  \label{fig:PCP_slide}
\end{figure*}
%%%%%%%%%%%%

Hence, (\ref{eqn:corrq}) can be rewritten as
\begin{align} \label{eqn:corr1} 
\check{p}[l'] \!&=\! \frac{1}{MN} \sum_{n=0}^{N\!-\!1} \Big| \sum_{l=0}^{M\!-\!1} \!\big(\! \sum_{\ell=0}^{L_{\rm{ch}}\!-\!1} h[\ell,nM+l] S[l-\theta-\ell,n] \nonumber
\end{align}
\begin{align}
&+ \zeta[nM\!+\!l] \big) \check{Z}^*[(l\!-\!l')_M,n] \Big| = \frac{1}{MN} \sum_{n=0}^{N\!-\!1} \Big| \sum_{l=l'_{\rm{p}}+l'}^{l''_{\rm{p}}+l'} \!\!\big(\!\! \sum_{\ell=0}^{L_{\rm{ch}}\!-\!1} \nonumber \\
& h[\ell,nM\!+\!l] S[l-\theta-\ell,n] \!+\! \zeta[nM+l] \big) \check{Z}^*[(l\!-\!l')_M,n] \Big|,
\end{align}
where the transmitted pilot in the delay-time domain $\check{Z}[l,n]$, represents the entries of $\check{\bZ}^q \,\, \forall q$. %, and can be obtained by
This pilot can be obtained by
\be \label{eqn:pilot_tx}
\check{Z}[(l-l')_M,\!n] \!=\!\!
\begin{cases}
\!S_{\rm{p}}[(l\!-\!l')_{\!M},\!n], &\!\!\!\!\! l'_{\rm{p}} \!\leq\!\! (l\!-\!l')_{\!M} \!\!\leq\! l''_{\rm{p}} \!\text{ and }\! l'_{\rm{p}} \!\leq\! l''_{\rm{p}}, \\
\!S_{\rm{p}}[(l\!-\!l')_{\!M},\!n], &\!\!\!\!\!\!\!\! \begin{array}[t]{l} l'_{\rm{p}} \!\leq\!\! (l\!-\!l')_{\!M} \!\text{ or }\! (l\!-\!l')_{\!M} \!\!\leq\! l''_{\rm{p}} \\ \!\text{ and }\! l'_{\rm{p}} \!>\! l''_{\rm{p}}, \end{array} \\
\!0, &\!\!\!\! \text{otherwise.}
\end{cases}
\ee
Furthermore, $S_{\rm{p}}[l,n]$ is the transmitted pilot in the delay-time domain and stacks the entries of $\check{\bO}^q \bF_N^{\rm{H}} \,\, \forall q$, with a transmit power of $\sigma^2_{\rm{p}}$.
Additionally, the transmitted signal $S[l,n]$ before CP addition, can be defined as
\be \label{eqn:sig_tx}
S[l,n] \!=\!\!
\begin{cases}\!
S_{\rm{p}}[l,n], & \!\!\text{ $l'_{\rm{p}} \!\leq\! l \! \leq \! l''_{\rm{p}}$,} \\
S_{\rm{d}}[l,n], & \!\!\text{otherwise,}
\end{cases}
\ee
for $S_{\rm{d}}[l,n]$ as the transmitted data symbols in the delay-time domain with a transmit power of $\sigma^2_{\rm{s}}$.
Since the transmitted pilot is nonzero only within the region $(l'_{\rm p} + l')_M \leq l \leq (l''_{\rm p} + l')_M$, the first term of equation~(\ref{eqn:corr1}) can be simplified accordingly to the second term.

Each PCP in the delay-time domain at a given time index, with length $2\check{L}_{\rm{p}}-1$, consists of two identical halves. These halves are formed by copying the last $\check{L}_{\rm{p}}-1$ samples of a ZC sequence as a CP, as illustrated in Fig.~\ref{fig:PCP_slide}. %Consequently,
Both halves correspond to the same $\check{L}_{\rm{p}}-1$ samples of a ZC sequence of length $\check{L}_{\rm{p}}$, hence, we refer to them as two distinct ZC sequences within the PCP.
Specifically, the CP in the PCP is called the first ZC sequence, and the identical data at the end of the PCP is considered the second ZC sequence. As shown in items (d–f) of Fig.~\ref{fig:PCP_slide}, the similarity between the two ZC sequences in the PCP and the complete PCP structure between the transmitted and received pilots across the delay dimension, results in three major peaks in the correlation function within the pilot region. To distinguish the major peaks from other insignificant peaks, an appropriate threshold must be selected to include the first major peak as the peak of interest. This first major peak occurs when the second ZC sequence in the sliding transmitted pilot is fully aligned with the first ZC sequence in the received pilot, this alignment is shown in Fig.~\ref{fig:PCP_slide}(d). In contrast, Fig.~\ref{fig:PCP_slide}(c) depicts the case one step before full alignment, where a one-sample misalignment between the two identical ZC sequences leads to a sharp drop in correlation amplitude. The same applies to the step immediately after full ZC sequence alignment. Several minor peaks also arise from another form of partial alignment in ZC sequences. Specifically, the last $\frac{\check{L}_{\rm{p}}+1}{2}$ samples of the second ZC sequence in the transmitted pilot may overlap with the first $\frac{\check{L}_{\rm{p}}+1}{2}$ samples of the first ZC sequence in the received pilot. This overlap results in a smaller, yet still distinct, correlation peak, as shown in Fig.~\ref{fig:PCP_slide}(b). Although this partial alignment does not produce a peak as strong as the main three, it still creates a local maximum that is higher than its surrounding values. Therefore, to robustly detect the first major peak in the pilot region, the detection threshold should be set between the amplitude of the minor peak caused by partial alignment as shown in Fig.~\ref{fig:PCP_slide}(b) and the amplitude of the first major peak resulting from full ZC sequence alignment as shown in Fig.~\ref{fig:PCP_slide}(d). This ensures that only the significant peaks, specifically the one of interest, are detected reliably.

%%%%%%%%%%%%
\begin{figure*}[!t]
  \centering 
  {\includegraphics[scale=0.35]{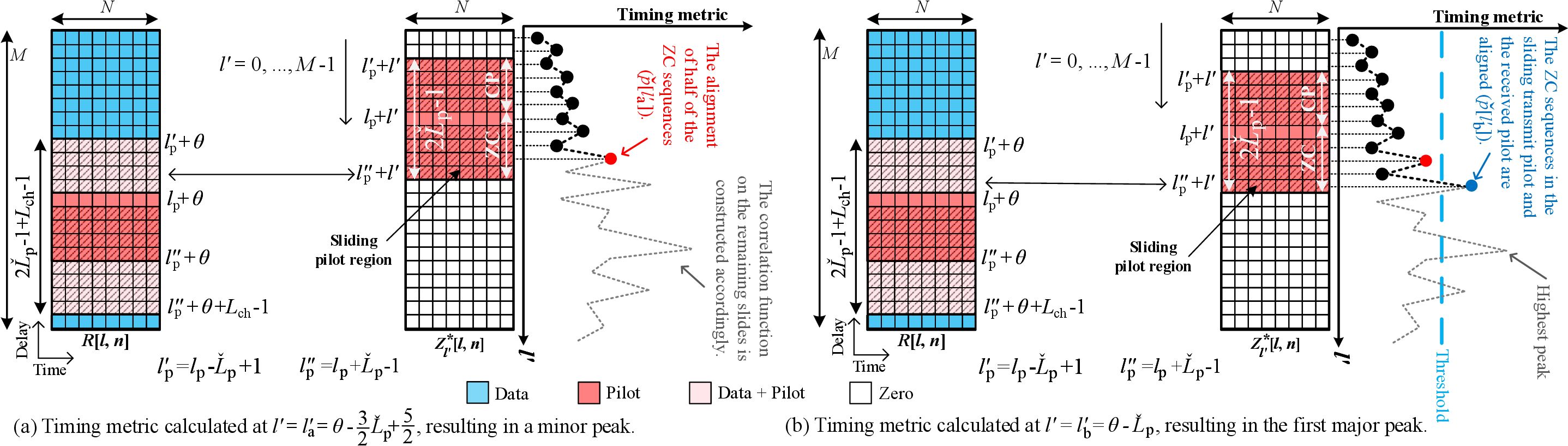}}
  \vspace{-0.3cm}
  \caption{The timing metric calculation in detail for a user after passing through a bandpass filter for $L_{\rm{ch}}=\check{L}_{\rm{p}}=5$.}
  \vspace{-0.4cm}
  \label{fig:pilot_slide}
\end{figure*}
%%%%%%%%%%%%

As shown in Fig.~\ref{fig:pilot_slide}, a sliding correlation is performed along the delay dimension between the transmitted pilot, which spans the rows $[l'_{\rm{p}}, l''_{\rm{p}}]$, and the received signal. Due to the effects of the channel and TO, the received pilot appears shifted and spans the range $[l'_{\rm{p}} + \theta, l''_{\rm{p}} + \theta + L_{\rm{ch}} - 1]$.
In this context, the peak generated by the partial alignment of half a ZC sequence is labeled $l'_{\rm{a}}$ in Fig.~\ref{fig:pilot_slide}(a).
Additionally, when the second ZC sequence in the sliding transmit pilot fully aligns with the first ZC sequence in the received pilot, the correlation function sharply increases and reaches the first major peak, labeled $l'_{\rm{b}}$ in Fig.~\ref{fig:pilot_slide}(b). As observed in this figure, this peak is located $\check{L}_{\rm{p}}$ samples before the highest peak, which corresponds to the full alignment of both the transmitted and received PCPs. 
The location of this first major peak can be used to identify the starting row of the pilot region in the delay-time domain. Therefore, to accurately detect the beginning of the received pilot region along the delay dimension, the threshold should be selected to lie between the correlation values at positions $l'_{\rm{a}}$ and $l'_{\rm{b}}$, i.e., $\check{p}[l'_{\rm{a}}] < \check{\mathcal{T}} < \check{p}[l'_{\rm{b}}]$.
As illustrated in Fig.~\ref{fig:pilot_slide}, the lower and upper bounds of the threshold occur at $l''_{\rm{p}}+l'=l_{\rm{p}}+\theta-1$ and $l''_{\rm{p}}+l'=l'_{\rm{p}}+\theta+\frac{\check{L}_{\rm{p}}+1}{2}$, respectively. These correspond to $l'=l'_{\rm{a}}=\theta-\frac{3}{2}\check{L}_{\rm{p}}+\frac{5}{2}$ and $l'=l'_{\rm{b}}=\theta-\check{L}_{\rm{p}}$.

We define (\ref{eqn:corr1}) as
\begin{align} \label{eqn:corr_sep} 
\check{p}[l'] = \frac{1}{MN} \sum_{n=0}^{N\!-\!1} \Big| P_{\rm{s}}[l',n] + P_{\zeta}[l',n] \Big|,
\end{align}
where $P_{\rm{s}}[l',n] = \sum_{l=l'_{\rm{p}}+l'}^{l''_{\rm{p}}+l'} \sum_{\ell=0}^{L_{\rm{ch}}\!-\!1} h[\ell,nM\!+\!l] S[l\!-\!\theta\!-\!\ell,n] \check{Z}^*[(l\!-\!l')_M,n]$ and
$P_{\rm{\zeta}}[l',n] = \sum_{l=l'_{\rm{p}}+l'}^{l''_{\rm{p}}+l'} \zeta[l,n] \check{Z}^*[(l-l')_M,n]$.
In the context of the triangle and reverse triangle inequalities, the following relations hold $|a|\!-\!|b| \!\leq\! |a\!+\!b| \!\leq\! |a| \!+\! |b|$. Thus, the threshold range can be written as 
$|p[l'_{\rm{a}}]| \!\leq\! \frac{1}{MN} \!\! \sum_{n=0}^{N\!-\!1} | P_{\rm{s}}[l'_{\rm{a}},n] | + | P_{\zeta}[l'_{\rm{a}},n] | \!<\! \check{\mathcal{T}} \!<\! \frac{1}{MN} \!\! \sum_{n=0}^{N\!-\!1} | P_{\rm{s}}[l'_{\rm{b}},n] | - | P_{\zeta}[l'_{\rm{b}},n] | \!\leq\! |p[l'_{\rm{b}}]|$.
As shown in Appendix~B, the threshold value should therefore be chosen within the range
\begin{align} \label{eqn:thr_final} 
& \frac{\sigma_{\rm{p}}}{M} \Big( \frac{\check{L}_{\rm{p}}+1}{2} \sigma_{\rm{p}} + \sqrt{\frac{3 (\check{L}_{\rm{p}}-1)}{2}} \sigma_{\rm{s}} + \sqrt{2 \check{L}_{\rm{p}}-1} \sigma_{\rm{\zeta}} \Big)
\, < \, \check{\mathcal{T}} \, < \, \nonumber \\
&\frac{ \sigma_{\rm{p}}}{M} \Big( (\check{L}_{\rm{p}}-1) \sigma_{\rm{p}} - \sqrt{\check{L}_{\rm{p}}} \sigma_{\rm{s}} - \sqrt{2\check{L}_{\rm{p}}-1} \sigma_{\rm{\zeta}} \Big).
\end{align}
This range can be normalized by the amplitude of the highest peak in the timing metric, which occurs at the slide where the transmitted PCP and received PCP are fully aligned. At this alignment, the expected value of the timing metric in (\ref{eqn:corrq}) is given by $\frac{2\check{L}_{\rm{p}} - 1}{M} \sigma^2_{\rm{p}}$. By neglecting the noise-related terms in the threshold range, the normalized range can be expressed as
\begin{align} \label{eqn:thr_final2} 
\frac{\check{L}_{\rm{p}}+1+\sqrt{6(\check{L}_{\rm{p}}\!-\!1)}\frac{\sigma_{\rm{s}}}{\sigma_{\rm{p}}}}{2(2\check{L}_{\rm{p}}-1)} \!<\! \mathcal{T} \!<\!
\frac{\check{L}_{\rm{p}}-1+\sqrt{\check{L}_{\rm{p}}}\frac{\sigma_{\rm{s}}}{\sigma_{\rm{p}}}}{2\check{L}_{\rm{p}}-1}.
\end{align}
Assuming $\frac{\sigma_{\rm{s}}}{\sigma_{\rm{p}}} \rightarrow 0$ and $\check{L}_{\rm{p}} \rightarrow \infty$, the threshold $\mathcal{T}$ lies within the range $0.25 \leq \mathcal{T} \leq 0.5$.

After estimating the TOs for different users, the pilot region for each user is identified. In the next stage, the CFOs of different users are estimated by $Q$ one-dimensional ML search problems as it will be explained in the next section. 

\vspace{-0.1cm}
\section{Proposed ML-Based CFO Estimation Technique} \label{sec:ML}
  \vspace{-0.05cm}
% %%%%%%%%%%%%%%%%%
Since both TO estimation techniques described above for different multiuser pilot structures, SU-PCP and MU-PCP, determine the location of each user's pilot, the CFO ML estimation problem can be formulated similarly for both scenarios, as the location of the received pilot signal for each user is known.
Hence, $L_{\rm{p}}$ in this section represents the pilot length for both multiuser pilot structures.

As the first step for CFO estimation, we stack the received pilot signal of user $q$ into a vector $\overline{\br}^q$.
Hence, the received samples after removing the CP from the pilot in the delay bins $\check{l}^q_{\rm{p}} = l^q_{\rm{p}} + \theta^q$ to $\check{l}^q_{\rm{p}} + L_{\rm{p}}-1$ over all the time slots are stacked in the vector $\overline{\br}^q\!=\![(\overline{\br}^q_0)^{\rm T}\!,\ldots\!,\!(\overline{\br}^q_{N\!-\!1})^{\rm T}]^{\rm{T}} \in \mathbb{C}^{NL_{\rm{p}}\times 1}$
where $\overline{\br}^q_n =[r^q[L_{\rm{cp}}+nM+\check{l}^q_{\rm{p}}], \ldots, r^q[L_{\rm{cp}}+nM+\check{l}^q_{\rm{p}}+L_{\rm{p}}-1]]^{\rm T}$.
It is worth noting that 
$l^q_{\rm{p}}=l_{\rm{p}} \,\, \forall q$ for MU-PCP, and
$\overline{\br}^q$ represents the received pilot region for user $q$, which can be located using either of the TO estimation techniques proposed in Sections (\ref{sec:TO1}) and (\ref{sec:TO2}).
Using (\ref{eqn:rec}), $\overline{\br}^q$ can be expressed as, 
\be \label{eqn:Mtx_rec1} 
\overline{\br}^q = \overline{\bPhi}(\varepsilon^q) \bOmega^q \overline{\s}^q + \overline{\boldsymbol{\eta}}^q, 
\ee
where 
$\overline{\s}^q$ represents the transmitted delay-time pilot for user $q$ as $\overline{\s}^q=[(\overline{\s}^q_0)^{\rm T},\ldots,(\overline{\s}^q_{N-1})^{\rm T}]^{\rm{T}} \in \mathbb{C}^{N L_{\rm{p}} \times 1}$, and $\overline{\s}^q_n$ denotes the pilot samples of user $q$ located in the delay bins from $L_{\rm{cp}}+l^q_{\rm{p}}$ to $L_{\rm{cp}}+l^q_{\rm{p}}+L_{\rm{p}}-1$ at time slot $n$.
Additionally, the CFO matrix in the pilot region is given by $\overline{\bPhi}(\varepsilon^q)\!=\!{\rm{diag}} [\overline{\bPhi}_0(\varepsilon^q),\ldots,\overline{\bPhi}_{N-1}(\varepsilon^q)] \in \mathbb{C}^{N{L}_{\rm{p}}}$ and $\overline{\bPhi}_n(\varepsilon^q)\!\!=\!
{\rm{diag}} \big[ e^{j \frac{2\pi \varepsilon^q (\check{L}_{\rm{cp}}\!+nM+\check{l}^q_{\rm{p}})}{N_{\rm{s}}}}\!\!\!,\ldots\!,e^{j \frac{2\pi \varepsilon^q (\check{L}_{\rm{cp}}\!+nM+\check{l}^q_{\rm{p}}\!+L_{\rm{p}}\!-\!1)}{N_{\rm{s}}}} \big]$.
The channel matrix for user $q$ in the pilot region that realizes the convolution operation can be represented by
${\bOmega}^q = {\rm{diag}}[\bOmega^q_0,\ldots,\bOmega^q_{N-1}] \in \mathbb{C}^{N L_{\rm{p}}}$ where $\bOmega^q_n$ has a structure that is shown in (\ref{eqn:ch2}), on the top of the next page.
Finally, $\overline{\boldsymbol{\eta}}^q \in \mathbb{C}^{NL_{\rm{p}} \times 1}$ is the noise vector in the pilot region of user $q$.
By interchanging the order of convolution in (\ref{eqn:Mtx_rec1}), the received pilot in the time slot $n$, can be expressed as
\begin{align}
\setcounter{equation}{23} \label{eqn:Mtx_rec2}
\overline{\br}^q_n &= \overline{\bPhi}_n(\varepsilon^q) \bA^q_n \bh^q_n + \overline{\boldsymbol{\eta}}^q_n,
\end{align}
where $\bA^q_n=[\bS^q_{n,0},...,\bS^q_{n,L_{\rm{p}}-1}]$, $\bS^q_{n,\ell}={\rm{diag}} [ (\!(\overline{\s}^q_n)\!)_{\ell} ] \in \mathbb{C}^{L_{\rm{p}}}$, 
$\bh^q_n=[(\bh^q_{n,0})^{\rm{T}},\ldots,(\bh^q_{n,L_{\rm{p}}-1})^{\rm{T}}]^{\rm{T}}$, and 
$\bh^q_{n,\ell}=[h^q[\ell, L_{\rm{cp}}+nM+\check{l}^q_{\rm{p}}],\ldots,h^q[\ell, L_{\rm{cp}}+nM+\check{l}^q_{\rm{p}}+L_{\rm{p}}-1]]^{\rm{T}}$ for $\ell=0,1,\ldots,L_{\rm{p}}-1$. Finally, $\overline{\boldsymbol{\eta}}^q_n \in \mathbb{C}^{L_{\rm{p}} \times 1}$ is the noise vector at time slot $n$. 

%%%%%%%%%%%%%%%%
\begin{figure*}[ht]
 \be
 \tag{23} \label{eqn:ch2}
 \begin{aligned}
 \bOmega^q_n \!=\!\!
 \begin{bmatrix}
 h^q[0,L_{\rm{cp}}+nM+\check{l}^q_{\rm{p}}] \!&\! 
 h^q[L_{\rm{p}}\!-\!1,L_{\rm{cp}}+nM+\check{l}^q_{\rm{p}}]  \!& \!\ldots\! &\!
 h^q[1,L_{\rm{cp}}+nM+\check{l}^q_{\rm{p}}]
\\
h^q[1,L_{\rm{cp}}+nM+\check{l}^q_{\rm{p}}+1] \!&\!
h^q[0,L_{\rm{cp}}+nM+\check{l}^q_{\rm{p}}+1] \!& \!\ldots\! &\!
h^q[2,L_{\rm{cp}}+nM+\check{l}^q_{\rm{p}}+1]
\\
\vdots &\vdots &\!\!\ddots\!\! &\vdots
\\
\!h^q[L_{\rm{p}}\!\!-\!1,L_{\rm{cp}}\!+\!nM+\check{l}^q_{\rm{p}}+L_{\rm{p}}\!\!-\!1] \!&\!
h^q[L_{\rm{p}}\!\!-\!2,L_{\rm{cp}}\!+\!nM+\check{l}^q_{\rm{p}}+L_{\rm{p}}\!\!-\!1] \!& \!\ldots\! &\!
h^q[0,L_{\rm{cp}}\!\!+\!nM+\check{l}^q_{\rm{p}}+L_{\rm{p}}\!\!-\!1]
\end{bmatrix}\!\!.
\end{aligned}
\ee
\vspace{-0.4cm}
\end{figure*}
%%%%%%%%%%%%%%%%

In high mobility scenarios, the channel coefficients fluctuate very rapidly. To facilitate accurate estimation in such scenarios, BEM models are used to capture the channel time variations. The CPF-BEM proposed in \cite{Muneer2015} for OFDMA systems outperforms other BEM models in terms of accuracy and simplicity.
Using CPF-BEM, the channel coefficients in (\ref{eqn:ch0}) can be expressed as
\be \label{eqn:ch_bem} h^q[\ell,\kappa]=\sum_{\gamma=0}^{\beta-1}B[\kappa',\gamma]c^q_{\ell}[\gamma], \ee
where $B[.,\gamma]$ is the CPF of degree $\gamma$, $\kappa'\!=\!\frac{2 \kappa-N_{\rm{s}}+1}{N_{\rm{s}}-1}$, and $c^q_{\ell}[\gamma]$ represents the basis coefficients for user $q$.
The basis functions can be obtained in a recursive manner using $B[\kappa',\gamma+1]=2 \kappa' B[\kappa',\gamma] - B[\kappa',\gamma-1]$ with the initial conditions $B[\kappa',0]=1$ and $B[\kappa',1]=\kappa'$.
The lower bound for the number of basis functions $\beta$, can be chosen based on $\beta \geq \lceil 2 \kappa_{\rm{max}} + 1 \rceil$.
Using (\ref{eqn:ch_bem}), $\bh^q_n$ can be approximated in terms of CPF coefficients, $ c^q_{\ell}[\gamma]$, as 
\be \label{eqn:bem1}
\bh^q_n=(\bI_{L_{\rm{p}}} \otimes \bB^q_n) \bc^q,
\ee
where $\bB^q_n \in \mathbb{C}^{L_{\rm{p}}^2 \times \beta L_{\rm{p}}}$ is the CPF-BEM matrix that stacks the values of $B[\kappa',\gamma]$ for $\kappa'\!=\!\frac{2 \kappa-N_{\rm{s}}+1}{N_{\rm{s}}-1}$, $\kappa \in \lbrace L_{\rm{cp}}\!+\!nM\!+\!\check{l}^q_{\rm{p}},\ldots, L_{\rm{cp}}\!+\!nM\!+\!\check{l}^q_{\rm{p}}\!+\!L_{\rm{p}}\!-\!1\rbrace$.
Additionally, $\bc^q=[(\bc^q_0)^{\rm{T}},\ldots,(\bc^q_{L_{\rm{p}}-1})^{\rm{T}}]^{\rm{T}} \in \mathbb{C}^{ \beta L_{\rm{p}} \times 1}$ for $\bc^q_{\ell}=[c^q_{\ell}(0),\ldots,c^q_{\ell}(\beta-1)]^{\rm{T}}$.
Substituting (\ref{eqn:bem1}) in (\ref{eqn:Mtx_rec2}), $\overline{\br}^q$ can be approximated by CPF as
\begin{align} \label{eqn:P_rec2} 
\overline{\br}^q &= \overline{\bPhi}(\varepsilon^q) \bG^q \bc^q + \overline{\boldsymbol{\eta}}^q,  
\end{align} 
where
$\bG^q\!=\![(\bG^q_0)^{\rm{T}}\!\!,\ldots,(\bG^q_{N\!-\!1})^{\rm{T}}]^{\rm{T}}$ and $\bG^q_n\!=\!\bA^q_n (\bI_{L_{\rm{p}}} \otimes \bB^q_n)$ for $n=0,1,\ldots,N-1$.

For a given set $(\bc^q,\varepsilon^q)$, the vector $\overline{\br}^q$ is assumed to have the Gaussian distribution with the mean 
$\overline{\bPhi}(\varepsilon^q) \bG^q \bc^q$ and covariance matrix $(\bm{\sigma}_\eta^q)^2 \bI_{NL_{\rm{p}}}$. Hence, the joint probability density function of $\overline{\br}^q$, parameterized by $(\tilde{\bc}^q,\tilde{\varepsilon}^q)$, is given by
\be \label{eqn:ML} 
f(\overline{\br}^q;\tilde{\bc}^q,\tilde{\varepsilon}^q) = \frac{1}{(\pi (\sigma^q_{\rm{\eta}})^2)^{NL_{\rm{p}}}} e^{-\frac{[\overline{\br}^q-\overline{\bPhi}(\tilde{\varepsilon}^q) \bG^q \tilde{\bc}^q]^{ \rm{H}} [\overline{\br}^q-\overline{\bPhi}(\tilde{\varepsilon}^q) \bG^q \tilde{\bc}^q]}{(\sigma^q_{\rm{\eta}})^2}}.
\ee
Thus, the ML estimates of the CPF coefficient vector and CFO are obtained as 
$(\hat{\bc}^q,\hat{\varepsilon}^q) = \arg \max_{\tilde{\bc}^q,\tilde{\varepsilon}^q}
\{ f(\overline{\br}^q;\tilde{\bc}^q,\tilde{\varepsilon}^q) \}$.
Taking the logarithm and removing the constant terms, the estimation problem can be simplified as 
$(\hat{\bc}^q,\hat{\varepsilon}^q) = \arg \max_{\tilde{\bc}^q,\tilde{\varepsilon}^q}
\{ g(\overline{\br}^q;\tilde{\bc}^q,\tilde{\varepsilon}^q) \}$,
where $g(\overline{\br}^q;\tilde{\bc}^q,\tilde{\varepsilon}^q)=\frac{-1}{(\sigma^q_{\rm{\eta}})^2} [\overline{\br}^q-\overline{\bPhi}(\tilde{\varepsilon}^q) \bG^q \tilde{\bc}^q]^{\rm{H}} [\overline{\br}^q-\overline{\bPhi}(\tilde{\varepsilon}^q) \bG^q \tilde{\bc}^q]$ is the joint cost function.
This maximization problem is solved by finding $\tilde{\bm{c}}^q$ that maximizes the joint cost function parameterized by
\be
\tilde{\bm{c}}^q(\tilde{\varepsilon}^q) = (\bG^q)^{\dag} \bPhi^{\rm{H}}(\tilde{\varepsilon}^q) \overline{\br}^q,
\label{eqn:ch} 
\ee
where $(\bG^q)^{\dag}=\big( (\bG^q)^{\rm{H}} \bG^q \big)^{-1} (\bG^q)^{\rm{H}}$.
Then, the obtained value of $\tilde{\bm{c}}^q$ is used to find a new cost function for $\tilde{\varepsilon}^q$ which
\begin{algorithm}[H] \label{alg:sync}
\caption{Proposed Synchronization Technique for MU-PCP}
\begin{algorithmic}[1]

\Statex \textbf{Input:} Received signal $\bm{r}$ and transmit pilots $\{\check{\bm{Z}}^q[l,n]\}_{q=0}^{Q\!-\!1}$

\State Calculate threshold $\mathcal{T}$ using (\ref{eqn:thr_final2})

\For{$q = 0$ to $Q-1$ \textbf{in parallel}}
    \State Apply bandpass filter $\bm{E}^q$ to $\bm{r}$ to obtain $\bm{r}^q$ using (\ref{eqn:BF})
    \State Convert $\bm{r}^q$ into 2D matrix form $\check{\bm{R}}^q[l,n]$
    \State Compute $\check{\bm{p}}^q [l']$ between $\check{\bm{R}}^q[l,n]$ and $\check{\bm{Z}}^q_{l'}[l,n]$ by (\ref{eqn:TO2_Corr})
    \State Detect peak group in $\check{\bm{p}}^q[l']$ above $\mathcal{T}$ using (\ref{eqn:TO_d})
    \State Estimate $\hat{\theta}^q$ using (\ref{eqn:TO_d2})
    \State Extract pilot region $\overline{\bm{r}}^q$ based on $\hat{\theta}^q$
    \State Construct CPF-BEM matrix $\bm{G}^q$ using (\ref{eqn:bem1}) and (\ref{eqn:P_rec2})
    \State Estimate $\hat{\varepsilon}^q$ via ML estimation by solving (\ref{eqn:cost2})
\EndFor

\Statex \textbf{Output:} Estimated TO and CFO, $\{\hat{\theta}^q, \hat{\varepsilon}^q\}_{q=0}^{Q-1}$

\end{algorithmic}
\end{algorithm}
\vspace{-0.3cm}
\noindent can be maximized by finding the CFO estimate through a search procedure. Thus, we fix $\tilde{\bvar}^q$ and $\tilde{\bm{c}}^q$ that maximizes $g(\tilde{\bm{c}}^q,\tilde{\varepsilon}^q)$ can be obtained as 
\be
g_{\rm{CFO}}(\tilde{\varepsilon}^q) = (\overline{\br}^q)^{\rm{H}} \bPhi (\tilde{\varepsilon}^q) \bG^q (\bG^q)^{\dag} \bPhi^{\rm{H}}(\tilde{\varepsilon}^q) \overline{\br}^q.
\label{eqn:cost} 
\ee
Next, the CFOs for all the users are estimated by solving multiple single-dimensional search problems centered around the zero as
\be \label{eqn:cost2} 
\hat{\varepsilon}^q={\arg} \max_{\tilde{\varepsilon}^q} \{g_{\rm{CFO}}(\tilde{\varepsilon}^q)\}. 
\ee

The proposed synchronization algorithm is summarized in Algorithm 1.
After CFO estimation, the CPF coefficients can be determined as $\hat{\bc}^q = (\bG^q)^{\dag} \bPhi^{\rm{H}} (\hat{\varepsilon}^q) \overline{\br}^q$.
Finally, the LTV channel in the delay-time domain is estimated using (\ref{eqn:ch_bem}).

\section{Spectral Efficiency and Complexity Analysis} \label{sec:BW}
In this section, we analyze the spectral efficiency and complexity of SU-PCP and MU-PCP structures. The first pilot structure inserts users' pilots into non-overlapping regions along the delay dimension. While this allocation strategy ensures interference-free pilot insertion, it impacts spectral efficiency, which can be evaluated using the metric
\begin{align} \label{eqn:BW1}
\lambda_{\rm{f}}^{\mathtt{SU-PCP}} = 
\frac{N(M-Q(2L_{\rm{ch}}-1))}{MN+L_{\rm{cp}}},
\end{align}
where $L_{\rm{p}}=L_{\rm{ch}}$, and SU-PCP includes full guards around the pilot along the Doppler dimension.
To prevent interference between users, based on the results in \cite{Raviteja2018}, a guard of at least $4 \kappa_{\rm{max}} = 4 \nu_{\rm{max}} NT$ samples is required along the Doppler dimension for the pilot region of each user. Thus, it is not necessary to set all symbols around the pilot to zero across the Doppler dimension, instead, placing $2\kappa_{\rm{max}}$ zero symbols on each side of the pilot is sufficient.
When partial zero guards, equivalent to the normalized Doppler spread, are used around each user's pilot in SU-PCP, (\ref{eqn:BW1}) converts to
\begin{align} \label{eqn:BW11}
\lambda_{\rm{p}}^{\mathtt{SU-PCP}} \!\!=\! \frac{MN -Q(2L_{\rm{ch}}\!\!-\!1)(4\kappa_{\rm{max}}\!\!+\!1)}{MN+L_{\rm{cp}}}\!.
\end{align}

However, the spectral efficiency of the MU-PCP structure can be determined using the metric 
\begin{align} \label{eqn:BW2}
\lambda^{\mathtt{MU-PCP}} =
% \frac{NT_{\rm{s}} \big(M-(2\check{L}_{\rm{p}}-1) \big)}{T_{\rm{s}} (MN+L_{\rm{cp}})} \!=\! 
\frac{N(M\!-\!(\beta+2 L_{\rm{ch}}-1))}{MN+L_{\rm{cp}}},
\end{align}
where $\check{L}_{\rm{p}}=\frac{\beta}{2}+L_{\rm{ch}}$.
To determine the conditions under which each proposed pilot structure performs better, $\lambda_0 = \frac{\lambda^{\mathtt{MU-PCP}}}{\lambda_{\rm{p}}^{\mathtt{SU-PCP}}} \geq 1$ and $\lambda_1 = \frac{\lambda^{\mathtt{MU-PCP}}}{\lambda_{\rm{f}}^{\mathtt{SU-PCP}}} \geq 1$ are defined, which can be simplified to $\beta \leq \big( \frac{Q}{N}(4 \kappa_{\rm{max}}+1)-1 \big)(2L_{\rm{ch}}-1)$ and $\beta \leq (Q-1)(2L_{\rm{ch}}-1)$, respectively.
As mentioned earlier $\beta \geq \lceil 2 \kappa_{\rm{max}} + 1 \rceil$ for a given user. Consequently, $\kappa_{\rm{max}} \leq 0.5 \frac{(2L_{\rm{ch}}-1)\frac{Q}{N}-2}{(2L_{\rm{ch}}-1)\frac{2Q}{N}+1}$ and $\kappa_{\rm{max}} \leq 0.5 \big((Q-1)(2L_{\rm{ch}}-1)-1 \big)$  are obtained as two conditions to ensure MU-PCP offers better spectral efficiency than SU-PCP with partial and full guards, respectively.
Based on the last equation for comparing MU-PCP and SU-PCP with full guards, the worst-case scenario for MU-PCP occurs when $Q=2$, leading to $\kappa_{\rm{max}} \leq L_{\rm{ch}}-1$.
This equation shows that as long as the normalized Doppler shift is smaller than the channel length, MU-PCP achieves better spectral efficiency than SU-PCP with full guards. Fig.~\ref{fig:BW5} compares the efficacy of the proposed pilot structures in multiuser scenarios where $Q > 1$.

%%%%%%%%%%%%
\renewcommand{\arraystretch}{1.4}
\begin{table}
\vspace{-0.2cm}
\centering
\caption{Computational Complexity}\vspace{-0.2cm}
\resizebox{0.489\textwidth}{!}
{\begin{tabular}{| c || c |}
\hline\hline
\textbf{Technique}
& \textbf{Number of CMs}
\\ \hline \hline
Estimation Technique for SU-OTFS
& $\frac{QN^2 L_{\rm{ch}}}{2} + MN(2L_{\rm{ch}}+Q-1)$
\\ \hline
Estimation Technique for MU-OTFS
& $\frac{Q N^2 (L_{\rm{ch}} \!+\! \kappa_{\rm{max}})}{2} \!\!+\!\! MN(2L_{\rm{ch}} \!+\! 2\kappa_{\rm{max}} \!+\! \log_2{\!MN} \!\!+\! Q \!-\! 1)$
\\ \hline
TO and Channel Estimation (Absorbed CFO)
&$MN(2L_{\rm{ch}}+Q-1)$
\\ \hline\hline
\end{tabular}
\label{Table:Complexity}
}
\vspace{-0.6cm}
\end{table}
%%%%%%%%%%%%
%%%%%%%%%%%
\begin{figure*}[!t]
\vspace{-0.5cm}
\begin{multicols}{3}
\centering 
  \includegraphics[scale=0.228]{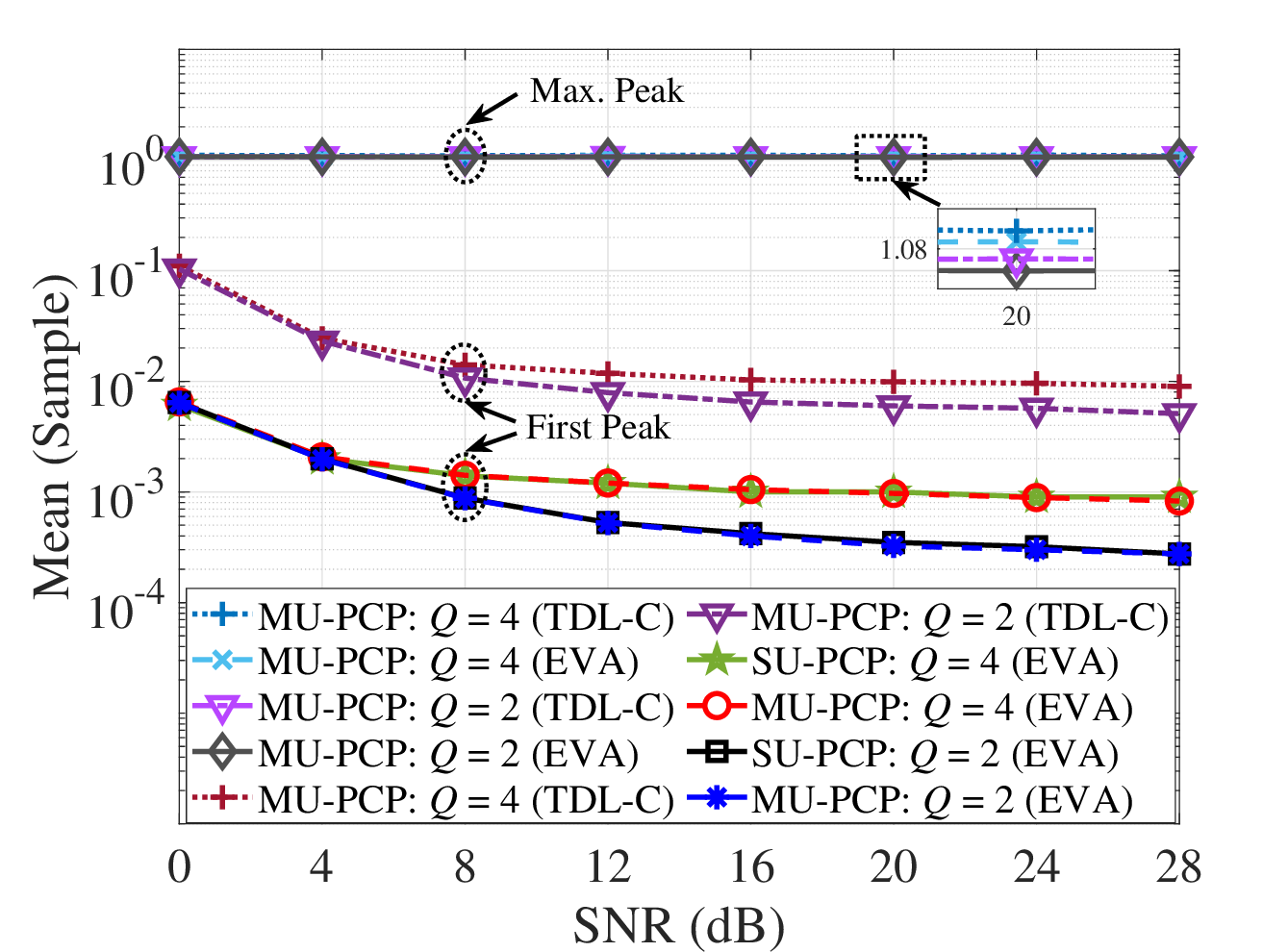}\par \vspace{-0.3cm} \caption{Performance of the proposed TO estimation techniques versus SNR at $\kappa_{\rm{max}} \approx 2.91$ for $16$-QAM.} \label{fig:TO_SNR}
    \includegraphics[scale=0.221]{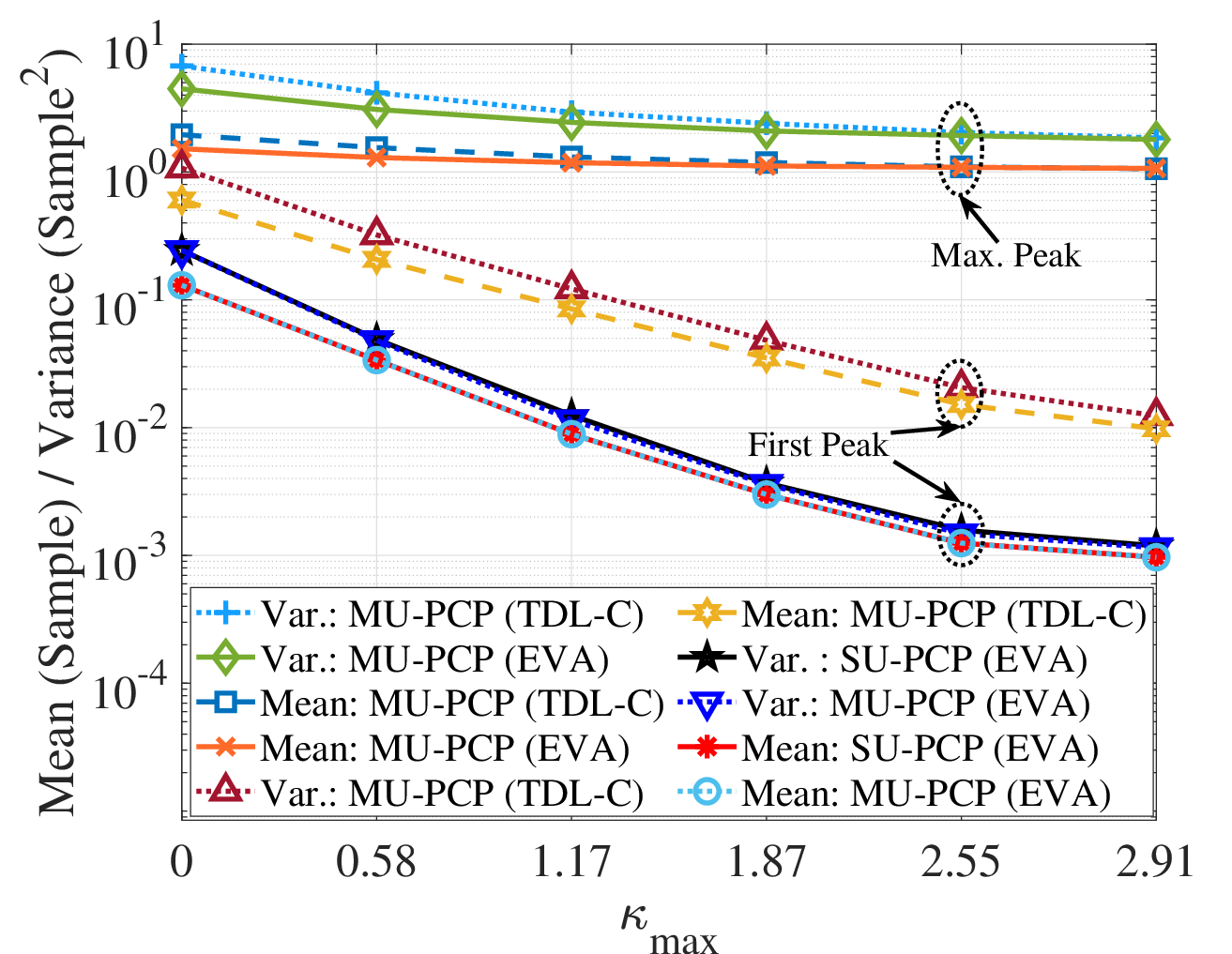}\par \vspace{-0.3cm} \caption{Performance of the proposed TO estimation techniques versus normalized maximum Doppler spread at ${\rm{SNR}}=20~{\rm{dB}}$ for $16$-QAM.} \label{fig:TO_Speed}
    \includegraphics[scale=0.228]{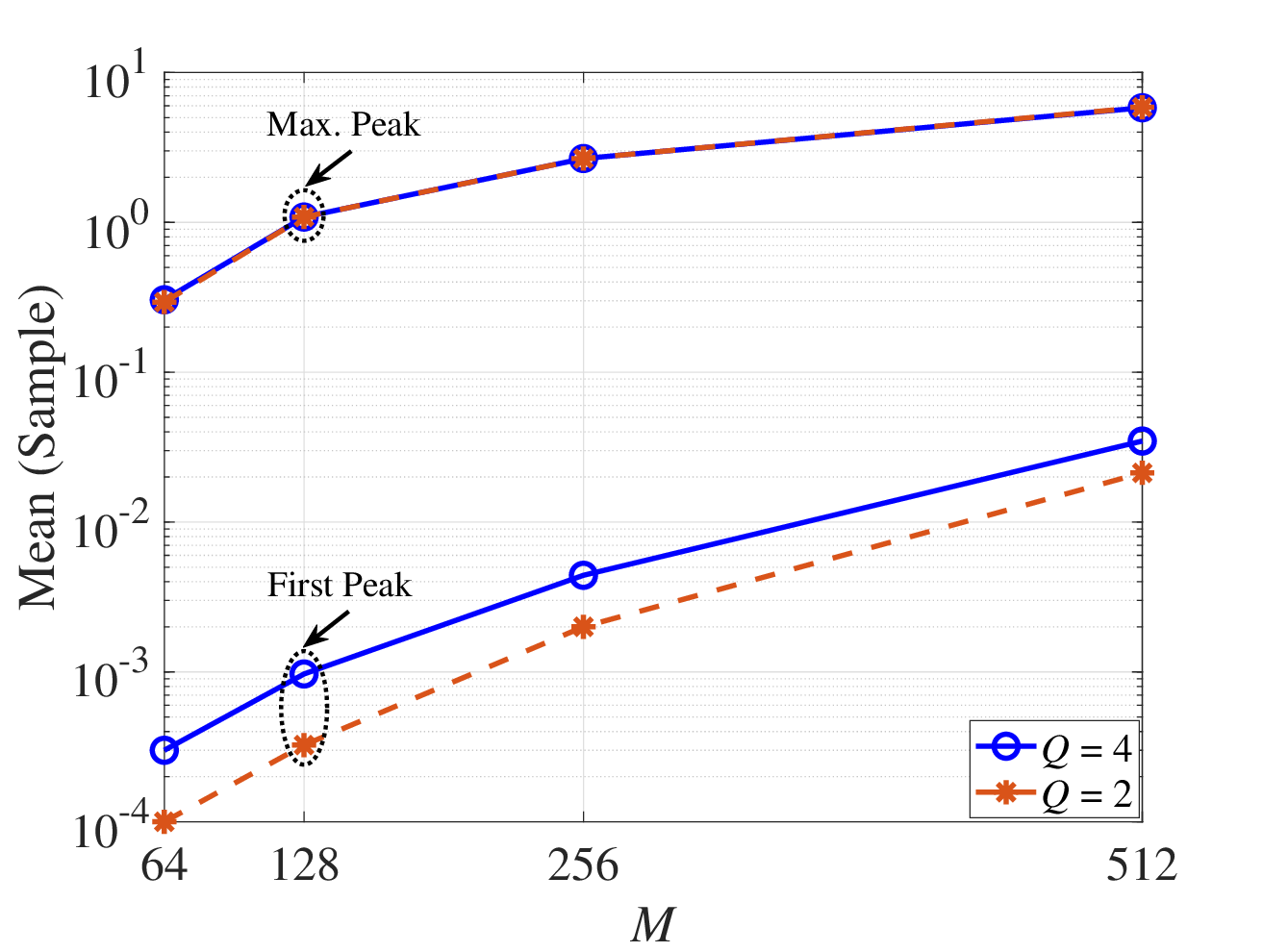}\par \vspace{-0.3cm} \caption{Performance of the proposed TO estimation technique for MU-PCP under varying bandwidths by adjusting the value of $M$.} \label{fig:BW_M1}
\end{multicols}\vspace{-0.6cm}
\end{figure*}
%%%%%%%%%%%

In terms of computational complexity, CFO estimation using ML requires $M^2 N^2$ complex multiplications (CMs). Additionally, channel estimation requires $MN$ CMs. Consequently, the total complexity of the ML solution for estimating CFOs and channel coefficients, whether using SU-PCP or MU-PCP, is $MN+M^2 N^2$ CMs. As shown in \cite{Bayat2023}, the ML implementation can be optimized to $\frac{N^2 L}{2}$ CMs, where the size of the pilot region is $(2L-1) \times N$. Since the proposed CFO estimation technique in this paper transforms the multiuser scenario into multiple single-user scenarios, this optimization can be applied here as well. Consequently, the proposed CFO and channel estimation technique requires $Q(MN+\frac{N^2 L}{2})$ CMs for the ML estimator used in both pilot structures. The correlation function complexity for TO estimation in both SU-PCP and MU-PCP requires $M(2L-1)N$ CMs. 
However, an additional filtering stage in MU-PCP adds an extra $MN\log_2{MN}$ CMs. Specifically, this filtering stage can be implemented using windowing in the frequency-Doppler domain. This process requires an $MN$-point fast Fourier transform (FFT) operation to transform the signal from the delay-time domain to the frequency-Doppler domain, followed by the application of a window that does not require any multiplication. Finally, an $MN$-point inverse FFT (IFFT) operation is performed to bring the filtered signal back to the delay-time domain.
Given that $L$ for SU-PCP and MU-PCP must be equal to $L_{\rm{ch}}$ and $\frac{\beta}{2}+L_{\rm{ch}}$, respectively, the total complexity of the proposed synchronization techniques is $Q(MN+\frac{N^2 L_{\rm{ch}}}{2})+2MNL_{\rm{ch}}-MN = \frac{QN^2 L_{\rm{ch}}}{2} + MN(2L_{\rm{ch}}+Q-1)$ for SU-PCP and $Q(MN+\frac{N^2 (L_{\rm{ch}}+\frac{\beta}{2})}{2})+2MN(L_{\rm{ch}}+\frac{\beta}{2})-MN+MN\log_2{MN} \approx \frac{Q N^2 (L_{\rm{ch}}+\kappa_{\rm{max}})}{2}+MN(2L_{\rm{ch}} + 2\kappa_{\rm{max}} + \log_2{MN} + Q-1)$ for MU-PCP.
Thus, while the synchronization technique for SU-PCP requires slightly lower processing overhead, it significantly degrades spectral efficiency, see Figs.~\ref{fig:BW5} and \ref{fig:comp5} in the Simulation Results Section.
To provide a better comparison, Fig.~\ref{fig:comp5} illustrates the ML-based channel estimation approach in which the CFO is absorbed into this stage. This approach requires $QMN$ CMs for channel estimation. Additionally, it involves $MN(2L_{\rm{ch}}-1)$ CMs for TO estimation to locate the pilot region for each user. Table~\ref{Table:Complexity} summarizes the computational complexity of the three discussed estimation techniques.

\vspace{-0.15cm}
\section{Simulation Results} \label{sec:Result}
\vspace{-0.05cm}
%%%%%%%%%%%%%%%%
This section numerically analyzes the performance of the proposed estimation techniques.
An MU-OTFS uplink scenario is considered where $M=128$ delay bins and $N=32$ Doppler bins are equally divided between $Q=2$ and $4$ users.
Unless otherwise stated, the extended vehicular A (EVA) channel model from \cite{3gpp} is used, with the channel length of $L^q_{\rm{ch}} \leq 10$ for $q=0,\ldots,Q-1$, a bandwidth of $\rm{BW}=3.84~\rm{MHz}$, and a carrier frequency of $f_{\rm{c}}=5.9~\rm{GHz}$.
The power of the PCP is set to $40~{\rm{dB}}$.
The SU-PCP structure shown in Fig.~\ref{fig:pilot0} spans over the delay dimension in $l^q_{\rm{p}}-{L}_{\rm{p}}+1,\ldots,l^q_{\rm{p}}+{L}_{\rm{p}}-1$, where $l^q_{\rm{p}} = l_{\rm{p}} + L_{\rm{p}} + q(2 {L}_{\rm{p}}-1)$ and $l_{\rm{p}}=\frac{M}{2} + \lfloor \frac{Q}{2} \rfloor - Q {L}_{\rm{p}}$, for ${L}_{\rm{p}}=10$ at $k_{\rm{p}}=\frac{N}{2}$.
The MU-PCP structure, as illustrated in Fig.~\ref{fig:pilot}, is deployed with $l_{\rm{p}}=M-1$ and $k^q_{\rm{p}} = \frac{1}{2} \lfloor \frac{N}{Q} \rfloor + q \lfloor \frac{N}{Q} \rfloor$ for $\check{L}_{\rm{p}}=10+\frac{\beta^q}{2}$.
For all the users $q$,  the order of the CPF-BEM is chosen as $1 \leq \beta^q \leq 12$ considering the normalized maximum Doppler spreads of $0 \leq \kappa_{\rm{max}} \le 2.91$.

%%%%%%%%%%%
\begin{figure*}[!t]
\vspace{-0.3cm}
\begin{multicols}{3}
\centering 
    \includegraphics[scale=0.234]{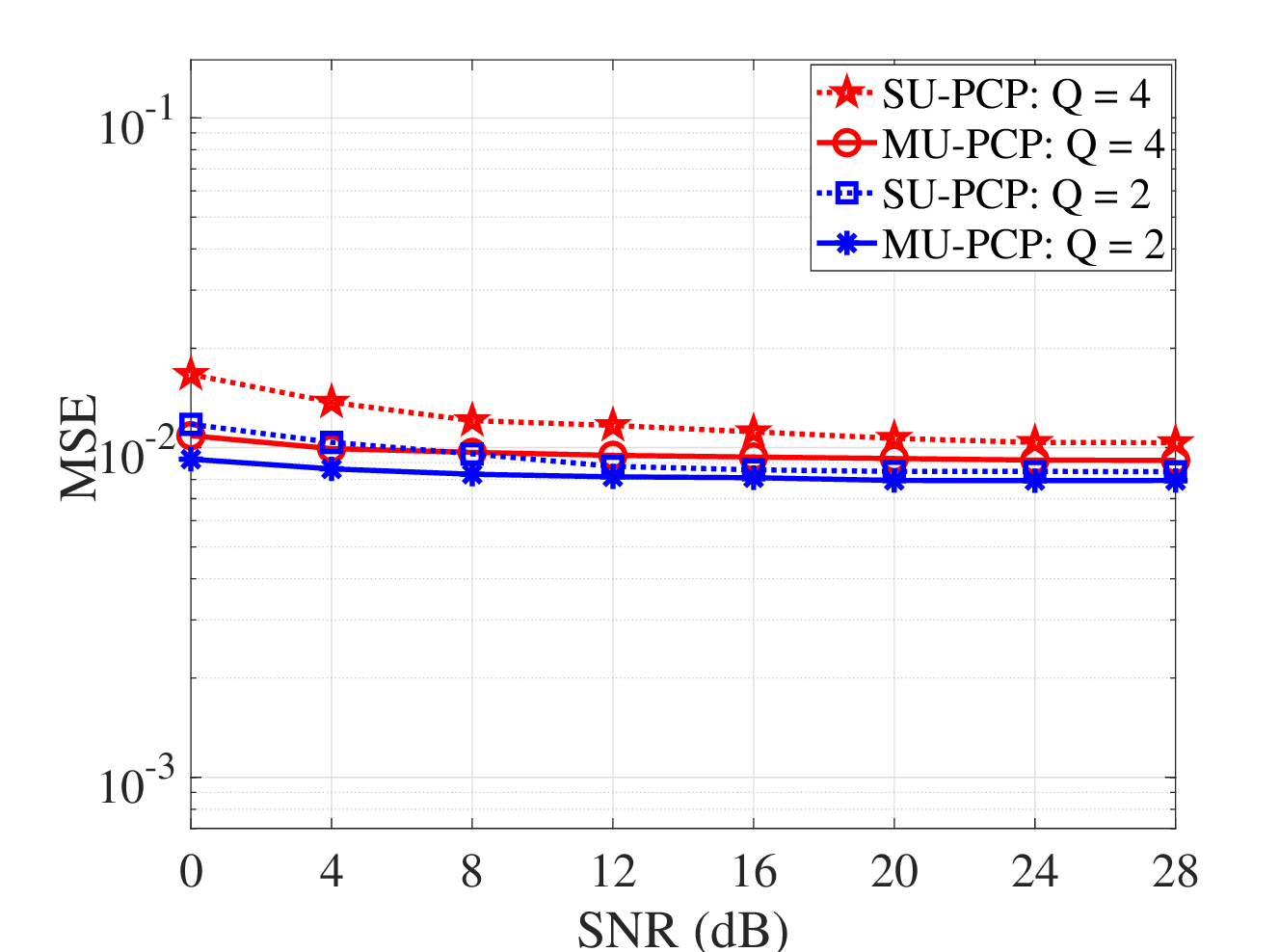}\par \vspace{-0.3cm} \caption{MSE performance of the proposed CFO estimation technique versus SNR at $\kappa_{\rm{max}} \approx 2.91$ for $16$-QAM.} \label{fig:CFO_SNR}
  \includegraphics[scale=0.228]{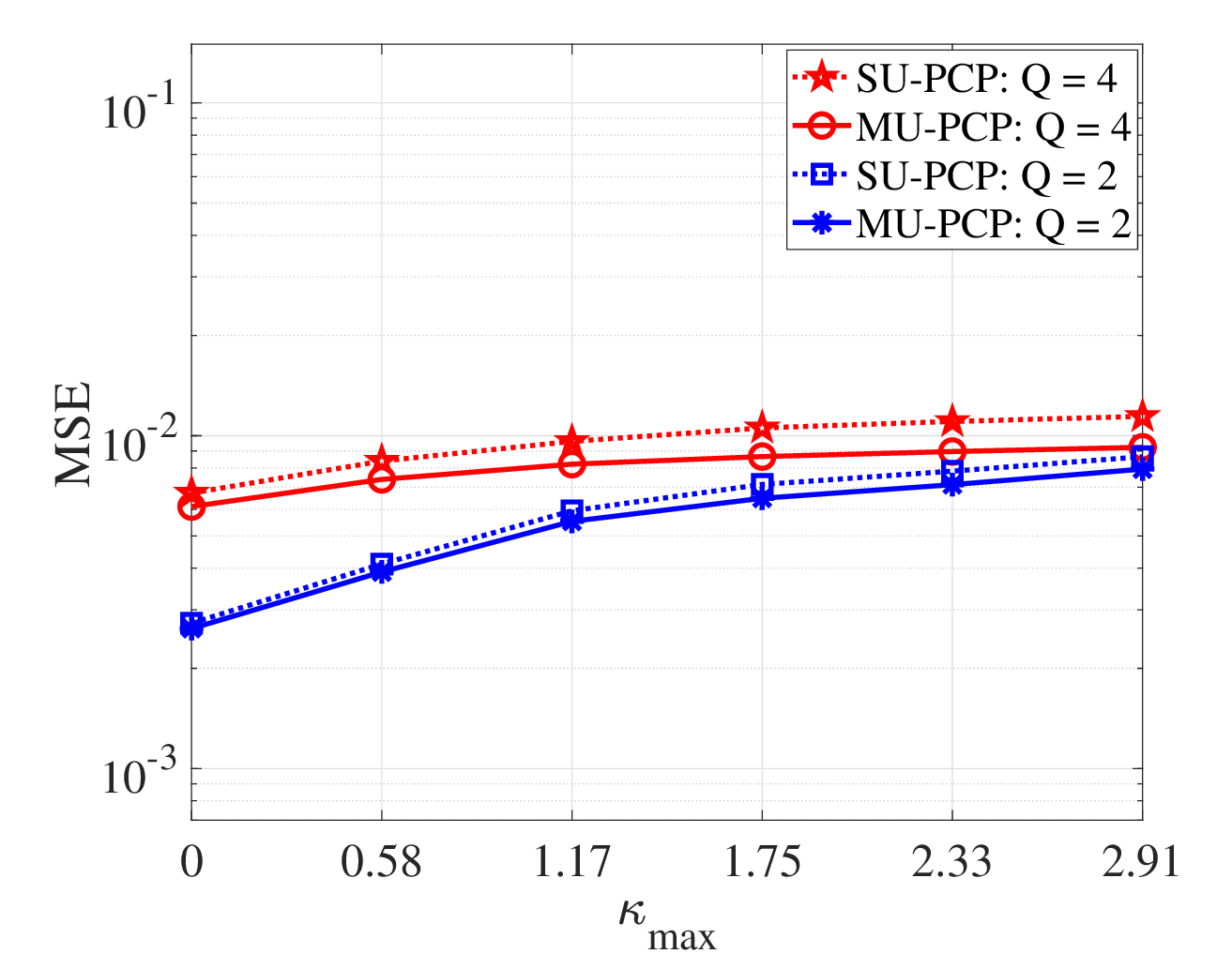}\par \vspace{-0.3cm} \caption{MSE performance of the proposed CFO estimation technique versus normalized maximum Doppler spread at ${\rm{SNR}}=20~{\rm{dB}}$ for $16$-QAM.} \label{fig:CFO_Speed}
    \includegraphics[scale=0.228]{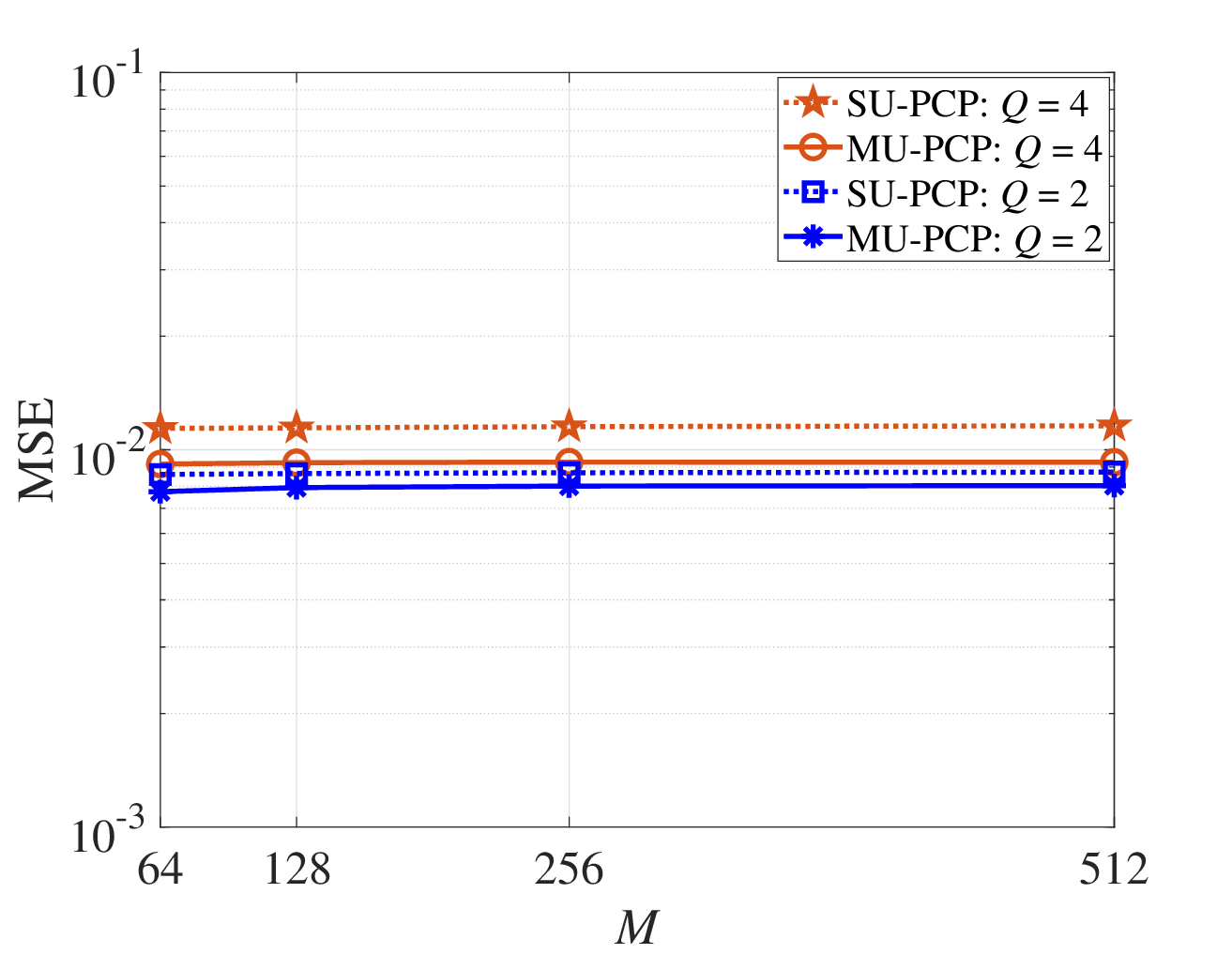}\par \vspace{-0.3cm} \caption{MSE performance of the proposed CFO estimation technique for varying bandwidths by changing $M$.} \label{fig:BW_M2}
\end{multicols}\vspace{-0.6cm}
\end{figure*}
%%%%%%%%%%%

Fig.~\ref{fig:TO_SNR} shows the mean of TO estimation error versus signal-to-noise ratio (SNR) for $Q\!=\!2,4$.
As shown in this figure, the proposed TO estimation techniques by finding the first major peak of the timing metric achieves orders of magnitude higher estimation accuracy than finding the maximum peak. These results indicate performance degradation for $Q\!=\!4$ compared to $Q\!=\!2$. This is due to the increased interference between the users' pilots as the guard between them shrinks by increasing $Q$ for fixed $M$ and $N$. 
This figure also shows that the proposed techniques, by selecting threshold values within the derived range, can achieve the same estimation accuracy. It is worth noting that both proposed TO estimation techniques, which rely on the first major peak, suffer from the fact that the ZC sequence is not completely independent from the data symbols, leading to several fake peaks in the correlation function when the LTV channel significantly deteriorates the received pilot.
Additionally, this figure presents the performance of the proposed TO estimation technique under the tapped delay line C (TDL-C) channel model \cite{3gpp_6}, where the first tap is not the strongest. In this scenario, although the performance degrades compared to the EVA channel, detecting the first major peak still yields better results than detecting the highest peak. It is worth noting that for the highest peak, the performance remains consistent across different channel models because (\ref{eqn:highest_peak}) adjusts the estimated TO using the channel delay mean.

In Fig.~\ref{fig:TO_Speed}, the mean and variance of the estimation error for the proposed TO estimation techniques are analyzed as a function of the normalized Doppler spread, i.e., $\kappa_{\rm{max}}$, for $Q\!=\!4$.
Based on these results, as Doppler spread increases, more accurate TO estimates are obtained which is due to the additional diversity gains provided by the time-selective channel. Similar to the previous figure, the proposed techniques that use the first major peak of the correlation function to estimate the TOs outperform those that rely on the maximum peak for TO estimation.
Additionally, both the proposed TO estimation techniques for SU-PCP and MU-PCP provide the same accuracy for different values of normalized Doppler spreads. The reason behind this similarity is that although in SU-PCP, the pilot regions for different users do not overlap along the delay dimension, due to the different values of TOs for different users, at the receiver the pilot regions will overlap, leading to interference among the pilot regions of different users. Thus, the amount of interference for this structure is similar to that for the MU-PCP.
This figure also shows the proposed TO estimation technique for MU-PCP under a TDL-C channel, utilizing both the highest and first major peaks. The resulting curves exhibit a similar trend to those observed under the EVA channel. %, with improvements attributed to the time-selectivity diversity.

Fig.~\ref{fig:BW_M1} shows the performance of the proposed TO estimation techniques under varying bandwidth conditions. Since the proposed estimation technique for both SU-PCP and MU-PCP yield nearly similar performance at $\rm{SNR}=20~\rm{dB}$, only the results for MU-PCP are presented in this figure. In this evaluation, the bandwidth is varied by increasing the number of delay bins, $M$. The Doppler spread caused by the time-varying channel, given by $\kappa_{\max}=\frac{f_{\rm{c}} v}{c \Delta \nu}$, is independent of $M$. Therefore, increasing the bandwidth by adjusting $M$ does not affect the time-selectivity diversity that the proposed TO estimation technique exploits. However, a larger $M$ results in a narrower threshold range, as described in (\ref{eqn:thr_final}). This reduces the gap between the intended minor and major peaks within the correlation function, making them harder to distinguish. As a result, this condition increases the timing estimation errors.

To prevent interference between the users, based on the results in \cite{Raviteja2018}, a guard of at least $4 \kappa_{\rm{max}}$ samples is required along the Doppler dimension for the pilot region of the user $q$.
However, for $N=32$, $Q=4$, and $\kappa_{\rm{max}}\approx2.91$, this condition is not satisfied and the CFO estimation performance is degraded by multiuser interference.
In Fig.~\ref{fig:CFO_SNR}, the mean squared error (MSE) performance of the proposed CFO estimation technique is evaluated as a function of SNR. The results demonstrate that the proposed technique achieves an MSE on the order of $10^{-2}$, and a higher accuracy for $Q=2$ as expected.
The location of the received pilot region varies across users due to different TO values. In the worst-case scenario for MU-PCP, when all users experience the same TO, their pilot regions fully overlap along the delay dimension, leading to significant MUI among pilot signals. In contrast, for SU-PCP, same TO values across users prevent pilot region overlap. However, for other TO values, SU-PCP also experiences MUI among pilot signals. As a result, the CFO estimation techniques for both pilot structures generally suffer from similar interference levels. Nevertheless, MU-PCP, with its longer pilot region of $\frac{\beta}{2}+L_{\rm{ch}}$, achieves higher CFO estimation accuracy.
In Fig.~\ref{fig:CFO_Speed}, the performance of the proposed CFO estimator is examined with respect to normalized Doppler spread. The results show that as the Doppler spread increases, the MSE performance deteriorates due to interference between the users' pilots.
%
%%%%
Fig.~\ref{fig:BW_M2} illustrates the performance of the proposed CFO estimation technique under varying bandwidth conditions for SU-PCP and MU-PCP. In this evaluation, the bandwidth is varied by increasing the number of delay bins, $M$.
In this comparison, the MSE remains nearly constant across different bandwidths. This is because the Doppler spread, which governs the time variation in the channel, remains unchanged. Consequently, the accuracy of the ML-based CFO estimator remains largely insensitive to variations in bandwidth resulting from changes in $M$.
%%%%%

%%%%%%%%%%%
\begin{figure*}[!t]
\vspace{-0.3cm}
\begin{multicols}{3}
\centering
    \includegraphics[scale=0.228]{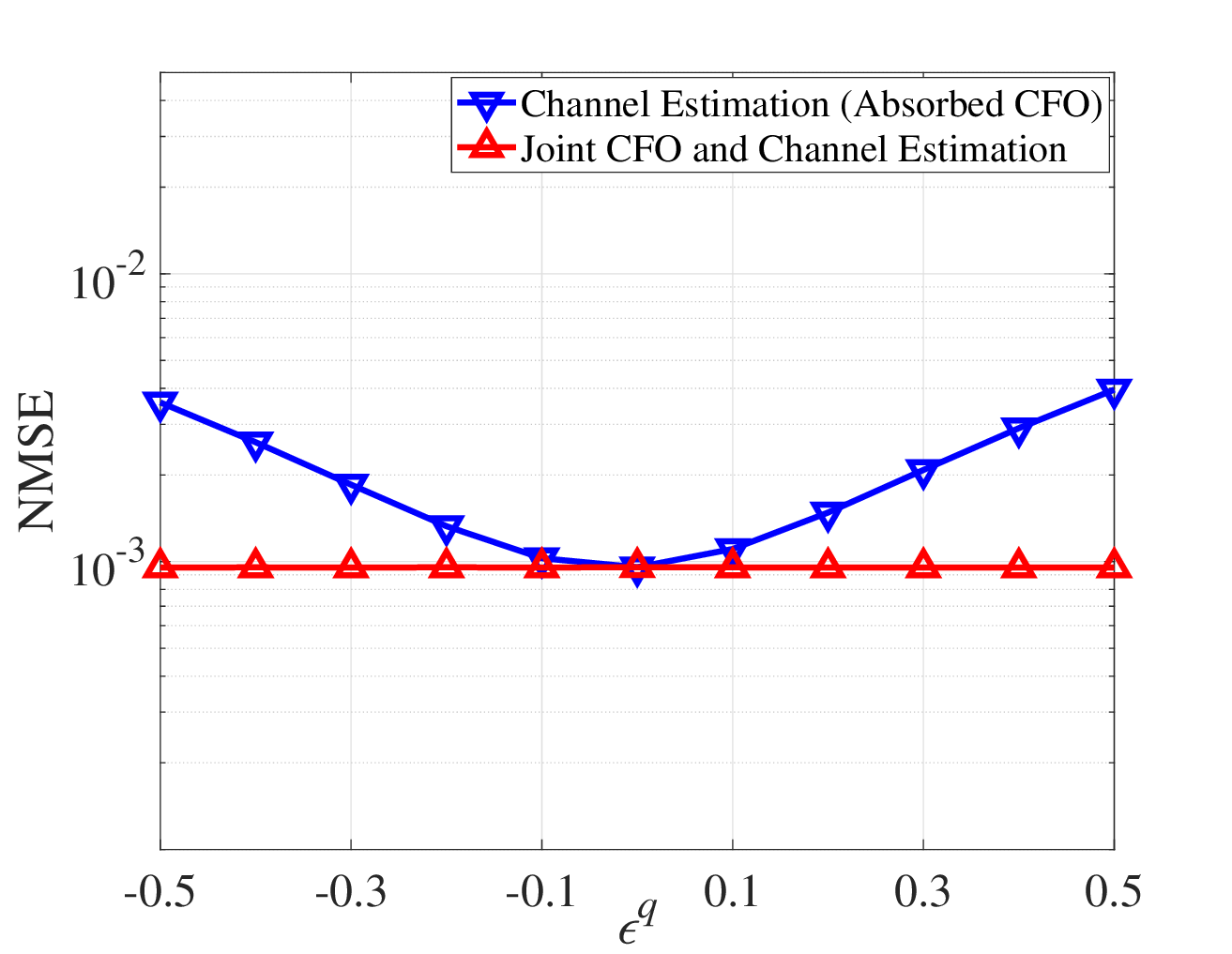}\par \vspace{-0.3cm} \caption{NMSE performance of the channel estimation technique at $\kappa_{\rm{max}} \approx 2.91$ and ${\rm{SNR}}=20~{\rm{dB}}$ for 16-QAM.} \label{fig:Ch_CFO}
   \includegraphics[scale=0.228]{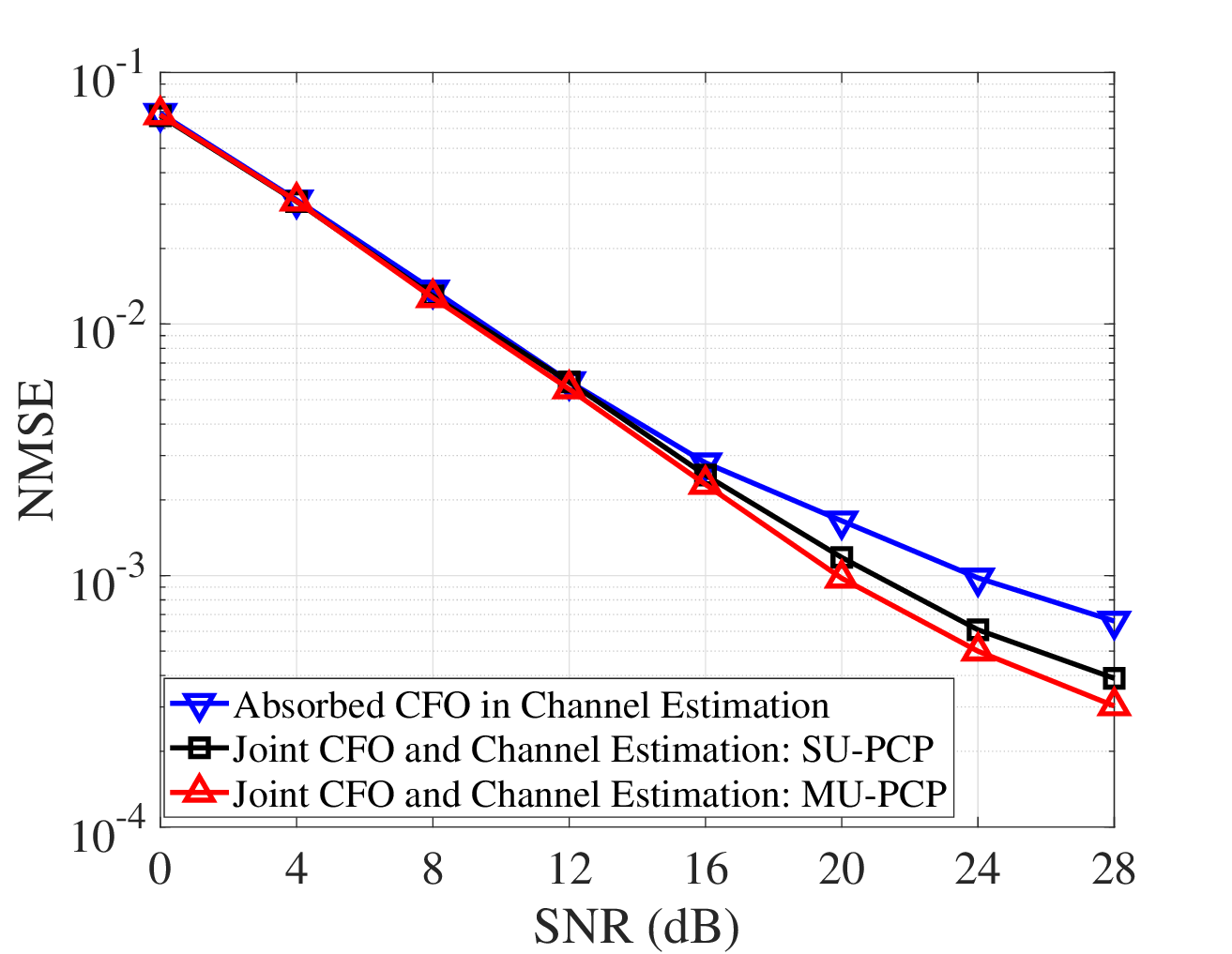}\par \vspace{-0.3cm} \caption{NMSE performance of the channel estimation for the different multiuser pilot structures at $\kappa_{\max}=2.91$ for $16$-QAM.} \label{fig:ch_snr}
    \includegraphics[scale=0.228]{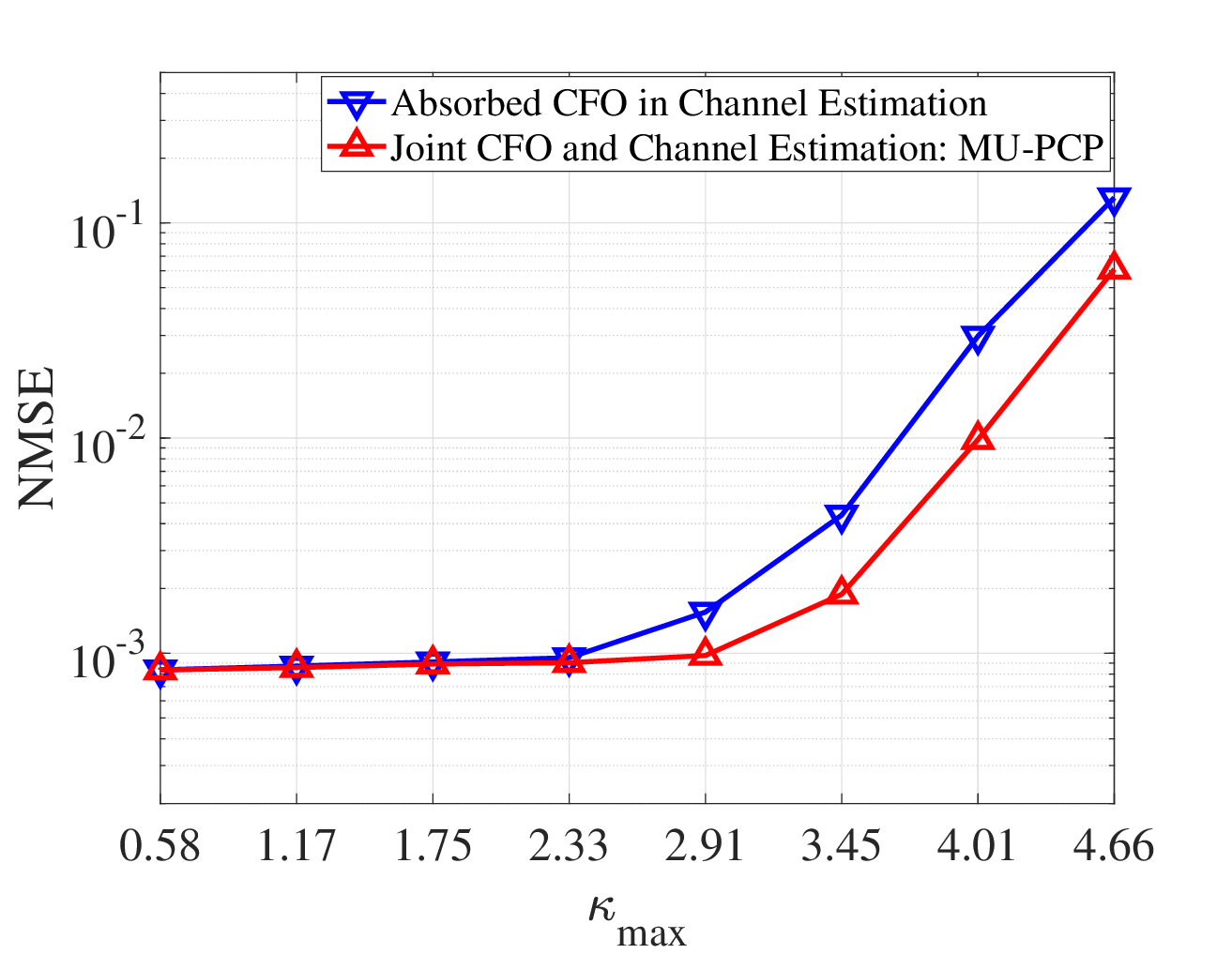}\par \vspace{-0.3cm} \caption{NMSE performance of channel estimation techniques versus normalized maximum Doppler spread at ${\rm{SNR}}=20~{\rm{dB}}$ with $16$-QAM.} \label{fig:ch_speed}
\end{multicols}\vspace{-0.6cm}
\end{figure*}
%%%%%%%%%%%

Fig.~\ref{fig:Ch_CFO} illustrates the sensitivity of channel estimation to normalized CFO, by showing the  normalized mean square error (NMSE) of channel estimation for different CFO values.
To focus purely on the effect of CFO on channel estimation, we assume perfect knowledge of TO in the results shown in this figure for both scenarios.
When the CFO is absorbed into the channel and it is estimated as a part of the channel, the estimation accuracy of the compound channel degrades as the CFO increases. In contrast, when the CFO is separately estimated more accurate channel estimate than the joint estimation case is obtained that is not sensitive to CFO.
In Fig.~\ref{fig:ch_snr}, the NMSE performance of the proposed joint channel and CFO estimation technique is presented for both SU-PCP and MU-PCP scenarios. The figure compares the proposed method with a baseline case in which the CFO is absorbed into the channel estimation. As shown, at high SNRs, the baseline case exhibits an error floor, which highlights the impact of absorbing the CFO into the channel estimation. This absorption amplifies the effect of the maximum Doppler spread in the time-varying channel, as discussed in Section~\ref{sec:System}.
Fig.~\ref{fig:ch_speed} shows that both approaches, absorbing the CFO into the channel estimation or compensating for it beforehand, offer similar performance up to a certain level of Doppler spread. However, as the Doppler spread increases, the channel estimation accuracy degrades for both methods, with the approach that absorbs the CFO experiencing more severe degradation.

\begin{figure*}[!t]
\vspace{-0.4cm}
\begin{multicols}{3}
\centering 
    \includegraphics[scale=0.228]{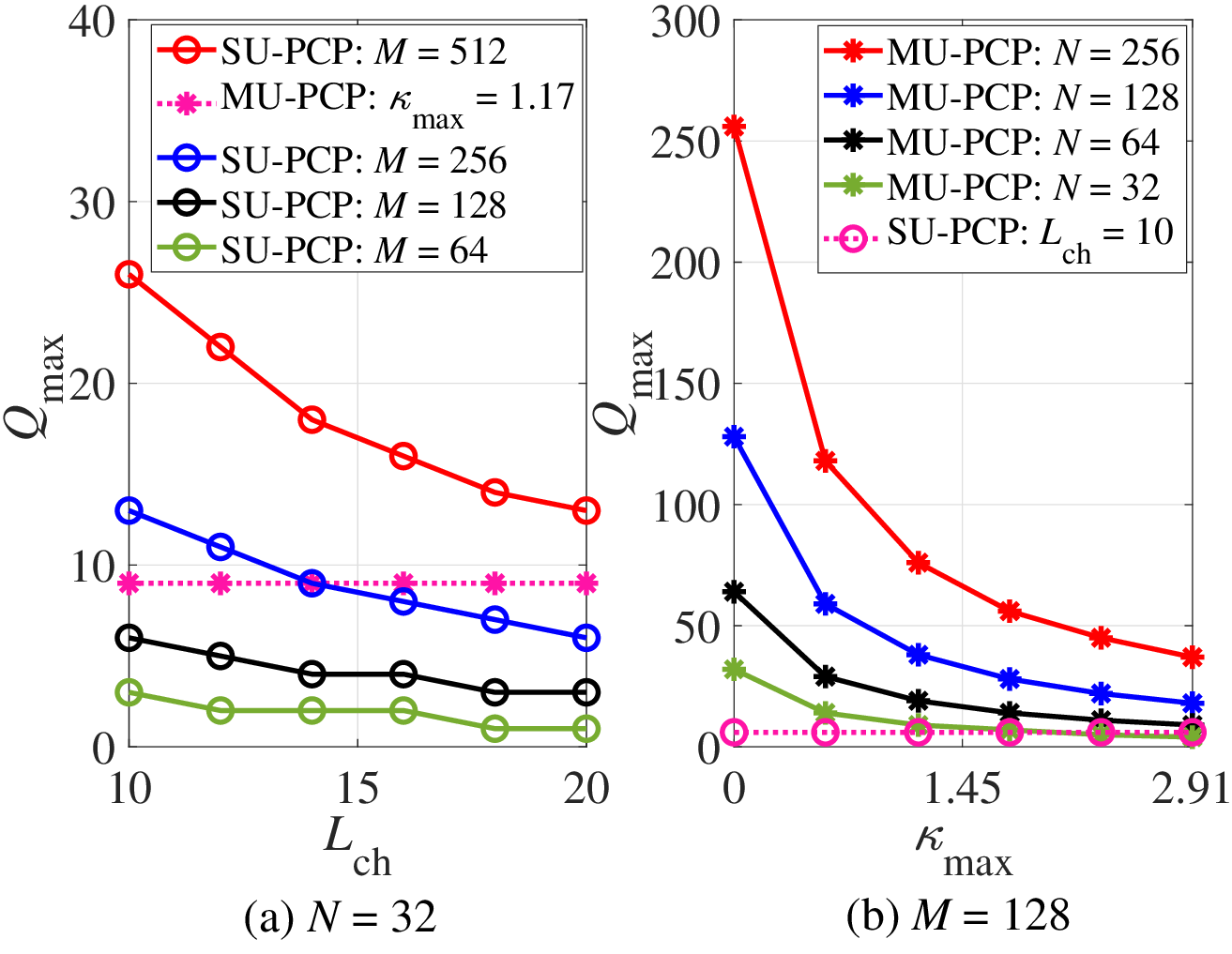}\par \vspace{-0.3cm} \caption{Number of allowed users versus (a) channel length for SU-PCP, and (b) normalized Doppler spread for MU-PCP.} \label{fig:Q}
    \includegraphics[scale=0.228]{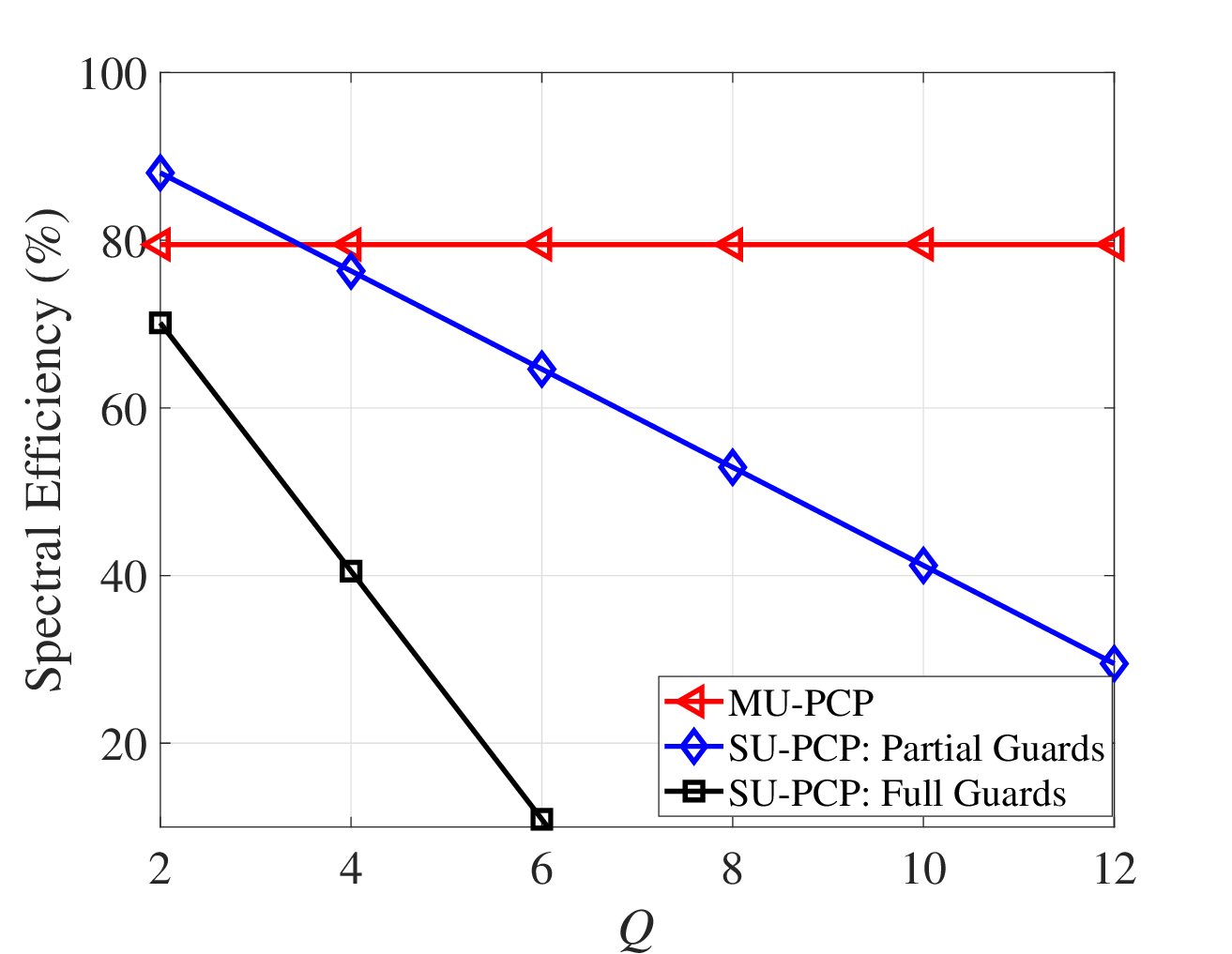}\par \vspace{-0.3cm} \caption{Spectral efficiency of the proposed MU-PCP and SU-PCP for different number of users at $\kappa_{\rm{max}} \approx 2.91$.} \label{fig:BW5}
    \includegraphics[scale=0.228]{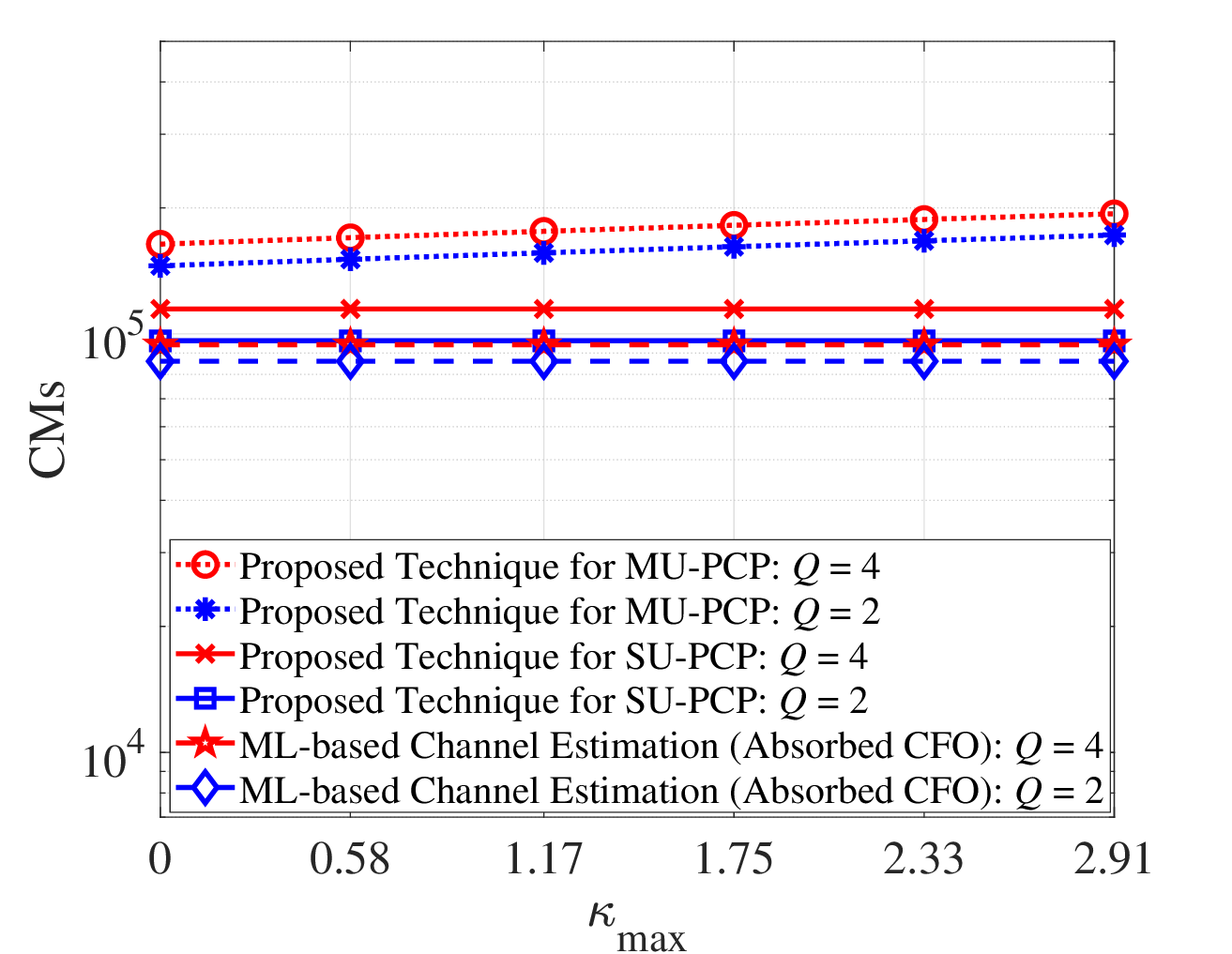}\par \vspace{-0.3cm} \caption{Complexity analysis of the proposed synchronization techniques for SU-PCP, MU-PCP, and ML-based channel estimation absorbing CFO.} \label{fig:comp5}
\end{multicols}\vspace{-0.6cm}
\end{figure*}

Fig.~\ref{fig:Q} illustrates the number of allowed users in both the SU-PCP and MU-PCP structures, based on the equations provided in Sections~\ref{sec:TO1} and~\ref{sec:TO2}. For SU-PCP, the maximum number of users that can be accommodated in an OTFS frame, assuming properly chosen ZC sequences that do not introduce limitations, is given by $Q_{\max}=\lfloor \frac{M}{2L_{\rm{ch}}-1} \rfloor$. As shown in Fig.~\ref{fig:Q}(a), increasing the channel length or decreasing the number of delay bins reduces the maximum number of users that can be supported.
Fig.~\ref{fig:Q}(b) illustrates the maximum number of users in MU-PCP, which is given by $Q_{\max}=\lfloor \frac{N}{4 \alpha \kappa_{\rm{max}}+1} \rfloor$. In this structure, increased time variation in the channel or reduced Doppler duration in the OTFS frame limits the maximum number of supported users. 
It is worth noting that in both figures related to SU-PCP and MU-PCP, the maximum number of users $Q_{\max}$ for the alternate pilot structure under certain parameters is also included for comparison.
For instance, in Fig.~\ref{fig:Q}(a), the SU-PCP curves are compared against a special case of MU-PCP. As expected from the expression for $Q_{\max}$ in MU-PCP, the number of users in this case remains constant despite increases in channel length. Conversely, in Fig.~\ref{fig:Q}(b), which shows the MU-PCP curves, the SU-PCP reference curve is also presented, which remains unaffected by changing $\kappa_{\max}$.

The spectral efficiency of the different multiuser pilot structures, namely MU-PCP and SU-PCP with partial and full guards, is compared in Fig.~\ref{fig:BW5}. This figure shows that a higher number of users significantly reduces the spectral efficiency of SU-PCP with full guards, while it has no effect on the spectral efficiency of MU-PCP. Additionally, the spectral efficiency of SU-PCP with partial guards is higher than that of MU-PCP when there are fewer than four users, however, as the number of users increases, its efficiency degrades significantly.
Fig.~\ref{fig:comp5} compares the required CMs for the synchronization techniques using two proposed pilot structures for uplink MU-OTFS in this paper. As shown in the figure, the synchronization technique using SU-PCP requires fewer CMs, as the one for MU-PCP necessitates filtering to separate the received pilot signals and a wider pilot region, i.e., $\frac{\beta}{2}+L_{\rm{ch}}$, for estimating both the CFOs and channel coefficients. This figure also shows that absorbing CFO estimation into the channel estimation slightly reduces the computational complexity. However, as discussed in Fig.~\ref{fig:ch_snr}, this reduction comes at the cost of lower estimation accuracy, i.e., up to $5~{\rm{dB}}$, in the presence of higher fractional CFOs.

\vspace{-0.2cm}
\section{Conclusion}\label{sec:Conclusion}
\vspace{-0.1cm}
%%%%%%%%%%%%
%%%%%%%%%%%
%%%%%%%%%%%
This paper, for the first time in literature, reported the challenges that arose in MU-OTFS uplink systems when the TOs and CFOs were absorbed into the channel. Specifically, it was demonstrated that accurate TO estimation was essential for locating the received pilots of different users, which was a prerequisite for CFO and channel estimation. Furthermore, the paper highlighted the importance of estimating the CFOs before channel estimation to achieve more accurate channel estimation performance than absorbing the CFO into the channel estimation stage. To address these challenges, a TO estimation technique was proposed for SU-PCP, which uses different ZC sequences, leveraging their zero-correlation property to estimate TOs. The second TO estimation technique, designed for MU-PCP, employed a bank of filters to separate the received signals of different users, and mitigate the interference caused by multiuser synchronization errors.
Using this bank of filters, the proposed correlation-based estimation technique utilized the periodicity in pilot signals to estimate multiple TOs. This step ensured that the users' pilot regions could be accurately found, addressing the misalignment of pilot regions caused by Doppler effects and propagation delays in the LTV channel.
A threshold range was derived to select the best possible threshold for finding the first major peak of the correlation function in the TO estimation techniques, ensuring highly accurate TO estimates.

Once the pilot region for each user was identified, the ML-based CFO estimation technique utilized the CPF-BEM to absorb the time variations of the LTV channel into the basis functions. This approach effectively reduced the multidimensional search for estimating multiple CFOs into multiple parallel one-dimensional ML search problems, significantly improving computational efficiency and estimation accuracy.
If each user's signal is synchronized separately, other users' signals will be misaligned. Thus, a model is needed that includes all users' channel effects and allows for joint TO and CFO compensation.
Hence, this paper formulated the received MU-OTFS uplink signal, leading to a derived input-output relationship and a compound channel matrix that includes the TOs, CFOs, and channels of all users. By jointly estimating and compensating all users’ TOs and CFOs, the proposed techniques ensured that the alignment of one user’s signal did not cause misalignment for other users.
The spectral efficiency of different pilot structures was mathematically analyzed. This analysis demonstrated that the proposed multiuser pilot structure provides better spectral efficiency than the existing multiuser pilot structure.
Finally, the proposed synchronization techniques were numerically analyzed, confirming their effective performances in addressing the synchronization challenges in MU-OTFS uplink transmissions under an LTV channel. The results demonstrated that the combination of a spectrally efficient pilot structure, advanced filtering, and robust estimation techniques provided a significant improvement in synchronization accuracy.

\vspace{-0.15cm}
\section*{Appendix A \\ Energy Concentration of LTV Channel}
\label{sec:apx_a}
\vspace{-0.0cm}

We consider a modified Jakes model such as the raised cosine Doppler spectrum, which offers a smoother approximation of realistic fading environments \cite{Teal2002}. The normalized
power spectral density (PSD) over the Doppler axis $\nu \in [-\nu_{\max}, \nu_{\max}]$ is given by $S(\nu) = \frac{1}{2 \nu_{\max}} \left[ 1 + \cos\left( \pi \nu / \nu_{\max} \right) \right]$ for $|\nu| \leq \nu_{\max}$, and zero otherwise.
This function is symmetric and integrates to unity, as $\int_{-\nu_{\max}}^{+\nu_{\max}} S(\nu) \, d\nu = 1$.
To compute the fraction of the total energy within a central portion of the Doppler support, e.g., $\alpha \nu_{\max}$ with $\alpha \in [0,1]$, we evaluate the normalized energy concentration within $[-\alpha \nu_{\max}, +\alpha \nu_{\max}]$ as
\be \label{eqn:apx_a2} \tag{35} E_{\rm{s}}(\alpha) = \int_{-\alpha \nu_{\max}}^{+\alpha \nu_{\max}} S(\nu) \, d\nu = \alpha + \dfrac{1}{\pi} \sin(\pi \alpha). \ee
Finally, to determine the value of $\alpha$ that contains $90\%$ of the total energy, we numerically solve $E_{\rm{s}}(\alpha) = \alpha + \dfrac{1}{\pi} \sin(\pi \alpha) = 0.9$. Evaluating for $\alpha = 0.4$ and $0.5$ gives $E_{\rm{s}}(0.4) \approx 0.89$ and $E_{\rm{s}}(0.5) \approx 0.91$.
Therefore, more than $90\%$ of the Doppler energy is concentrated within $\alpha = 0.5$, corresponding to $50\%$ of the total Doppler span.

\vspace{-0.2cm}
\section*{Appendix B \\ TO Threshold Calculation} \label{sec:apx_b}
\vspace{-0.2cm}
To determine the upper bound of the threshold $\check{\mathcal{T}}$ using $\check{p}[l'_{\rm{b}}]$, each component of the timing metric at $l'_{\rm{b}}$ is evaluated.
Assuming $l'_{\rm{p}} < l''_{\rm{p}}$, the signal term is first calculated as
\begin{align} \label{eqn:corr_sig_sig} 
P_{\rm{s}}[l'_{\rm{b}},n] \!=\!\!\!\!\! \sum_{l=l'_{\rm{p}}+l'_{\rm{b}}}^{l''_{\rm{p}}+l'_{\rm{b}}} \! \sum_{\ell=0}^{L_{\rm{ch}}\!-\!1} \!\! h[\ell,nM\!+\!l] S[l\!-\!\theta\!-\!\ell,n] S^*_{\rm{p}}[(l\!-\!l'_{\rm{b}})_M,n]. \vspace{-0.2cm}
\end{align}
As shown in Fig.~\ref{fig:pilot_slide}(b), in this equation, the $\check{L}_{\rm{p}}$ rows of the sliding transmit pilot overlap with the signal portion of the received signal in the range $l'_{\rm{p}} + l'_{\rm{b}} \leq l \leq l_{\rm{p}} + l'_{\rm{b}}$, and $\check{L}_{\rm{p}} - 1$ rows with the pilot portion of the received signal in the range $l_{\rm{p}} + l'_{\rm{b}} + 1 \leq l \leq l''_{\rm{p}} + l'_{\rm{b}}$.
Thus, (\ref{eqn:corr_sig_sig}) can be decomposed into two terms as follows
\begin{align} \label{eqn:corr_sig_pilot} 
&
P_{\rm{s}}[l'_{\rm{b}},n] \!=\!\!\!\!\!\!\! \sum_{l=l'_{\rm{p}}+l'_{\rm{b}}}^{l_{\rm{p}}+l'_{\rm{b}}} \! \sum_{\ell=0}^{L_{\rm{ch}}\!-\!1} \!\! h[\ell,\!nM\!\!+\!l] S_{\rm{d}}[l\!-\!\theta\!-\!\ell,n] S^*_{\rm{p}}[(l\!-\!l'_{\rm{b}})_{\!M},n] \nonumber \\
&\!\!+ \!\!\!\!\!\!\! \sum_{l=l_{\rm{p}}+l'_{\rm{b}}+1}^{l''_{\rm{p}}+l'_{\rm{b}}} \! \sum_{\ell=0}^{L_{\rm{ch}}\!-\!1} \!\! h[\ell,nM\!\!+\!l] S_{\rm{p}}[l\!-\!\theta\!-\!\ell,n] S^*_{\rm{p}}[(l\!-\!l'_{\rm{b}})_{\!M},n].
\end{align}
Recalling $l'_{\rm{b}}=\theta-\check{L}_{\rm{p}}$, $l'_{\rm{p}}=l_{\rm{p}}-\check{L}_{\rm{p}}+1$, and $l''_{\rm{p}}=l_{\rm{p}}+\check{L}_{\rm{p}}-1$, equation (\ref{eqn:corr_sig_pilot}) can be written as
\begin{align} \label{eqn:corr_sig_pilot2} 
&P_{\rm{s}}[\theta\!-\!\check{L}_{\rm{p}},n] \!=\! \Big( \!\sum_{l=l_{\rm{p}}+\theta-2\check{L}_{\rm{p}}+1}^{l_{\rm{p}}+\theta-\check{L}_{\rm{p}}} \sum_{\ell=0}^{L_{\rm{ch}}\!-\!1} \! h[\ell,nM\!+\!l] S_{\rm{d}}[l\!-\!\theta\!-\!\ell,n] \nonumber
\end{align}
\begin{align}
& S_{\rm{p}}^*[(l-\theta+\check{L}_{\rm{p}})_M,n] \Big) \,+\, \Big( \, \sum_{l=l_{\rm{p}}+\theta-\check{L}_{\rm{p}}+1}^{l_{\rm{p}}+\theta-1} \, \sum_{\ell=0}^{L_{\rm{ch}}-1}  h[\ell,nM+l] \nonumber
\end{align}
\begin{align}
% &
S_{\rm{p}}[l-\theta-\ell,n] S_{\rm{p}}^*[(l-\theta+\check{L}_{\rm{p}})_M,n] \Big).
\end{align}
The first term corresponds to the correlation between the pilot signal and the data symbols, both of length $\check{L}_{\rm{p}}$, across all time indices affected by the channel.
Additionally, the second term in this equation represents the overlap of two similar ZC sequences, namely the second ZC sequence in the transmitted pilot, $S_{\rm{p}}[(l-\theta+\check{L}_{\rm{p}})_M,n]$, and the first ZC sequence in the received pilot, $S_{\rm{p}}[l-\theta-\ell,n]$. Each sequence has a length of $\check{L}_{\rm{p}} - 1$, and the overlap occurs across all time indices affected by the channel.
Given the variances of the signal, pilot, and noise as $\sigma^2_{\rm{s}}$, $\sigma^2_{\rm{p}}$, and $\sigma^2_{\zeta}$, respectively, the expected amplitude of the timing metric component related to the signal in (\ref{eqn:corr_sep}) at $l'_{\rm{b}}$ can be calculated as follows
\begin{align} \label{eqn:corr_th_sig_upper} 
\mathbb{E} \{ |P_{\rm{s}}[l'_{\rm{b}},n]| \} &\!\geq\! \big( (\check{L}_{\rm{p}}-1) \sigma^2_{\rm{p}} - \sqrt{\check{L}_{\rm{p}}} \sigma_{\rm{s}} \sigma_{\rm{p}} \big) \sum_{\ell=0}^{L_{\rm{ch}}\!-\!1} \! \rho_{\rm{ch}}[\ell],
\end{align}
where $\rho_{\rm{ch}}[\ell]$ represents the power delay profile of the channel for $\ell=0,\ldots,L_{\rm{ch}}-1$ and recalling triangle inequality as $|a|-|b| \leq |a+b| \leq |a|+|b|$ for $a,b$ arbitrary complex values.

In the following step, the noise term in (\ref{eqn:corr_sep}) is expanded as %follows
\begin{align} \label{eqn:corr_noise_pilot}
&P_{\rm{\zeta}}[l'_{\rm{b}},n] = \sum_{l=l'_{\rm{p}}+l'_{\rm{b}}}^{l''_{\rm{p}}+l'_{\rm{b}}} \zeta[l,n] S_{\rm{p}}^*[(l\!-\!l'_{\rm{b}})_M,n] \nonumber \\
&= \sum_{l=l_{\rm{p}}+\theta-2\check{L}_{\rm{p}}+1}^{l_{\rm{p}}+\theta-1} \!\! \zeta[l,n] S_{\rm{p}}^*[(l\!-\!\theta\!+\!\check{L}_{\rm{p}})_M,n],
\end{align}
that represents the multiplication of the AWGN and transmitted pilot in the delay-time domain.
The expected value for the amplitude of the timing metric related to the noise in (\ref{eqn:corr_sep}) at $l'_{\rm{b}}$ can be calculated as $\mathbb{E} \{ |P_{\rm{\zeta}}[l'_{\rm{b}},n]| \} =  \sqrt{2\check{L}_{\rm{p}}-1} \sigma_{\rm{\zeta}} \sigma_{\rm{p}}$.
While the nature of the AWGN and pilot are independent of the location of the signals in the expected value, $\mathbb{E} \big\{ |P_{\rm{\zeta}}[l'_{\rm{a}},n]| \big\} \!=\! \mathbb{E} \big\{ |P_{\rm{\zeta}}[l'_{\rm{b}},n]| \big\}$.
Finally, the upper bound of the threshold is obtained as
\begin{align} \label{eqn:corr_th_upper} 
&\mathbb{E} \big\{ |\check{p}[l'_{\rm{b}}]| \big\} \geq \frac{1}{MN} \sum_{n=0}^{N-1} \mathbb{E} \big\{ |P_{\rm{s}}[l'_{\rm{b}},n] | - |P_{\rm{\zeta}}[l'_{\rm{b}},n] | \big\} \geq \frac{\sigma_{\rm{p}}}{M}
\nonumber \\ 
& \Big( \big( (\check{L}_{\rm{p}}\!-\!1) \sigma_{\rm{p}} - \sqrt{\check{L}_{\rm{p}}} \sigma_{\rm{s}} \big) \!\! \sum_{\ell=0}^{L_{\rm{ch}}\!-\!1} \! \rho_{\rm{ch}}[\ell] \!-\! \sqrt{2\check{L}_{\rm{p}}\!-\!1} \sigma_{\rm{\zeta}} \Big).
\end{align}

%%%%%%%%%%%%%%%%%%%%%%%%%%%%%%%%%%%%%%%%%%%%%
Using (\ref{eqn:corr_sig_pilot2}), the signal-related term at the lower bound of the threshold $\check{\mathcal{T}}$, corresponding to $l' = l'_{\rm{a}}$, can be derived. As shown in Fig.~\ref{fig:pilot_slide}(a), at this point, $\frac{\check{L}_{\rm{p}}+1}{2}$ samples of the second ZC sequence in the sliding transmit pilot cover the beginning of the first ZC sequence in the received pilot region. The remaining rows of the sliding transmit pilot overlap only with the signal portion of the received signal. 
Recalling $l'_{\rm{a}}=\theta-\frac{3}{2} \check{L}_{\rm{p}}+\frac{5}{2}$, this partial pilot alignment can be obtained by
\begin{align} \label{eqn:corr_sig1_pilot} 
&P_{\rm{s}}[l'_{\rm{a}},n] \!=\!\!\!\!\! \sum_{l=l'_{\rm{p}}+l'_{\rm{a}}}^{l''_{\rm{p}}+l'_{\rm{a}}} \!\! \sum_{\ell=0}^{L_{\rm{ch}}\!-\!1} \!\! h[\ell,nM\!+\!l] S_{\rm{d}}[l\!-\!\theta\!-\!\ell,n] S^*_{\rm{p}}[(l\!-\!l'_{\rm{a}})_{\!M},n] \nonumber \\
&= \!\!\!\!\! \sum_{l=l_{\rm{p}}+\theta-\frac{5}{2}\check{L}_{\rm{p}}+\frac{7}{2}}^{l_{\rm{p}}+\theta-\check{L}_{\rm{p}}+1} \!\! \sum_{\ell=0}^{L_{\rm{ch}}\!-\!1} \!\! h[\ell,nM\!+\!l] S_{\rm{d}}[l\!-\!\theta\!-\!\ell,n] S_{\rm{p}}^*[(l\!-\!\theta\!+\!\frac{3}{2}\check{L}_{\rm{p}} \nonumber \\
& -\frac{5}{2})_M,n] + \sum_{l=l_{\rm{p}}+\theta-\check{L}_{\rm{p}}+2}^{l_{\rm{p}}+\theta-\frac{1}{2}\check{L}_{\rm{p}}+\frac{3}{2}} \sum_{\ell=0}^{L_{\rm{ch}}-1} h[\ell,nM+l] S_{\rm{p}}[l-\theta-\ell,n] \nonumber \\
& S^*_{\rm{p}}[(l-\theta+\frac{3}{2}\check{L}_{\rm{p}}-\frac{5}{2})_M,n].
\end{align}
The expected value of the first term can be calculated as
$\sqrt{\frac{3(\check{L}_{\rm{p}}\!-\!1)}{2}} \sigma_{\rm{s}} \sigma_{\rm{p}} \sum_{\ell=0}^{L_{\rm{ch}}-1} \!\! \rho_{\rm{ch}}[\ell]$.
The second term involves two segments of ZC sequences, each of length $\frac{\check{L}_{\rm{p}}+1}{2}$, that lead to the expected value of $\frac{\check{L}_{\rm{p}}+1}{2} \sigma^2_{\rm{p}} \sum_{\ell=0}^{L_{\rm{ch}}-1} \rho_{\rm{ch}}[\ell]$.
Thus, the lower bound of the threshold can be calculated as
\begin{align} \label{eqn:corr_th_sig_lower} 
&\mathbb{E} \big\{ |p[l'_{\rm{a}},n]| \big\} \leq \frac{1}{MN} \sum_{n=0}^{N-1} \mathbb{E} \big\{ |P_{\rm{s}}[l'_{\rm{a}},n] | + |P_{\rm{\zeta}}[l'_{\rm{a}},n] | \big\} 
\nonumber \\ &= \frac{\sigma_{\rm{p}}}{M} \big( (\sqrt{\frac{3 (\check{L}_{\rm{p}}\!-\!1)}{2}} \sigma_{\rm{s}} \!+\! \frac{\check{L}_{\rm{p}}\!+\!1}{2} \sigma_{\rm{p}}) \sum_{\ell=0}^{L_{\rm{ch}}-1} \!\! \rho_{\rm{ch}}[\ell] \!+\! \sqrt{2 \check{L}_{\rm{p}}\!-\!1} \sigma_{\rm{\zeta}} \big).
\end{align}
Finally, assuming that $\sum_{\ell=0}^{L_{\rm{ch}}-1} \rho_{\rm{ch}}[\ell]=1$, the threshold range can be defined as 
\begin{align} \label{eqn:thr_final0} 
& \frac{\sigma_{\rm{p}}}{M} \Big( \sqrt{\frac{3 (\check{L}_{\rm{p}}-1)}{2}} \sigma_{\rm{s}} + \frac{\check{L}_{\rm{p}}+1}{2} \sigma_{\rm{p}} + \sqrt{2 \check{L}_{\rm{p}}-1} \sigma_{\rm{\zeta}} \Big)
\, < \, \check{\mathcal{T}} \, < \, \nonumber \\
&\frac{ \sigma_{\rm{p}}}{M} \Big( (\check{L}_{\rm{p}}-1) \sigma_{\rm{p}} - \sqrt{\check{L}_{\rm{p}}} \sigma_{\rm{s}} - \sqrt{2\check{L}_{\rm{p}}-1} \sigma_{\rm{\zeta}} \Big).
\end{align}
\vspace{-0.4cm}

%%%%%%%%%%%%%%%%%%%%%%%%%%%%%%%%%%%%%%%%
\bibliographystyle{IEEEtran} 
\bibliography{IEEEabrv,references}

\end{document}